\documentclass[preprint,12pt,sort&compress]{elsarticle}
\usepackage{amsmath,amssymb,booktabs,graphicx,float,caption,array}
\usepackage{tikz}
\usepackage{url}
\usepackage{xcolor}
\journal{Journal of Computational Physics}

\newcommand{\NEW}[1]{#1}
\newcommand{\NEWMATH}[1]{\ensuremath{\displaystyle #1}}
\newcommand{\meanstd}[4]{\shortstack{\ensuremath{#1\times10^{#2}}\\[-1pt]\textcolor{black!65}{\ensuremath{\scriptstyle\pm #3\times10^{#4}}}}}
\newcommand{\bestmeanstd}[4]{\shortstack{\ensuremath{\mathbf{#1}\times10^{#2}}\\[-1pt]\textcolor{black!65}{\ensuremath{\scriptstyle\pm #3\times10^{#4}}}}}

\begin{document}
\begin{frontmatter}
\title{A Compensated Koopman Neural Operator with Selective State-Space Dynamics for Unsteady Flows}

\makeatletter
\renewcommand{\corref}[1]{\def\@corref{$\ast$\hskip-1pt}}
\makeatother
\author[aff1]{Tangying Lv}
\author[aff1]{Yuanjun Dai\corref{cor1}}
\author[aff2]{Zhenxu Sun\corref{cor1}}
\affiliation[aff1]{organization={School of Mathematics and Computer Sciences, Nanchang University},
             city={Nanchang},
             postcode={330031},
             country={China}}
\affiliation[aff2]{organization={Key Laboratory for Mechanics in Fluid Solid Coupling Systems, Institute of Mechanics, Chinese Academy of Sciences},
             city={Beijing},
             postcode={100190},
             country={China}}

\begin{abstract}
Stable prediction of unsteady flows requires accurate multiscale spatial representation and robust temporal propagation. We introduce the Compensated Koopman U-shaped Neural Operator (CoKo-UNO), which combines a U-shaped spectral backbone with Koopman-dominated latent propagation. Finite-dimensional Koopman truncation produces a state-dependent residual that is repeatedly reinjected during autoregressive rollout. CoKo-UNO models this residual with a selective state-space model (SSM), a principled input-dependent compensation mechanism, together with resolution-adaptive compensatory skip connections and an overlapping-warmup rollout strategy. \NEW{Across four benchmark problems, CoKo-UNO achieves the lowest mean rollout error among all compared methods. Its largest gain is a $76.76\%$ reduction relative to the strongest baseline, while requiring about $41.40\%$ of RNO's training time.} These results show that explicit residual compensation improves stable autoregressive prediction of unsteady flows.
\end{abstract}

\end{frontmatter}

\clearpage
\section{Introduction}
\begingroup
\renewcommand{\thefootnote}{\fnsymbol{footnote}}
\footnotetext[1]{\raggedright Corresponding authors.\\ \hspace*{1.8em}\textit{Email addresses:} \texttt{iankin.dai@ncu.edu.cn} (Yuanjun Dai), \texttt{sunzhenxu@imech.ac.cn} (Zhenxu Sun).}
\endgroup
The prediction of unsteady flows relies almost entirely on the solution of time-dependent partial differential equations (PDEs), which encode the understanding of the physical world accumulated by mathematicians and physicists over the past centuries \cite{courant1962methods}. To solve these equations accurately, generations of researchers have constructed numerical schemes that represent spatial structure and advance solutions in time with ever higher fidelity; classical discretizations such as the essentially non-oscillatory and weighted essentially non-oscillatory schemes \cite{harten1987uniformly,shu1988efficient} remain landmarks of computational physics. Nevertheless, high-fidelity numerical simulation remains expensive, since accuracy demands fine meshes and small time steps. The cost is particularly severe for strongly nonlinear systems such as the Navier--Stokes equations, where repeated simulation under varying flow conditions is often the computational bottleneck.

Recent progress in artificial intelligence for PDEs offers a complementary route. Physics-informed neural networks (PINNs) incorporate governing equations into the training loss and can solve forward and inverse problems without a mesh \cite{raissi2019physics,karniadakis2021physics}, but they usually approximate the solution for one specific set of initial and boundary conditions and require retraining when these change. Neural operators instead learn mappings between function spaces and therefore provide reusable surrogates for parameterized PDE families \cite{kovachki2023neural,lu2021learning}. Despite this promise, two challenges remain central for the prediction of unsteady flows: spatial representation, i.e., how to faithfully represent multiscale flow structures; temporal propagation, i.e., how to advance the solution stably under autoregressive rollout, since a strong spatial backbone alone does not control the accumulation of temporal errors.

On the spatial side, two families of neural operators are representative. Deep Operator Networks (DeepONets) combine a branch network that encodes the input field with a trunk network that encodes query coordinates, providing a general pointwise parameterization of operators \cite{lu2021learning}. The Fourier Neural Operator (FNO) instead performs nonlocal integral operations in Fourier space, which has proved particularly effective for learning global interactions of flow fields \cite{li2021fourier}. Subsequent work has extended the spectral formulation in several directions: geometry-aware and graph-based variants handle complex geometries and unstructured meshes \cite{li2022geometry,li2023gino,mno2020,pfaff2021learning}, attention- and convolution-based architectures improve the expressiveness of the learned kernel \cite{hao2023gnot,cao2021choose,brandstetter2022message,raonic2023cno}, and physics-informed training incorporates PDE constraints into operator learning \cite{pino2024}. Orthogonal improvements concern discretization handling, latent representations, pretraining, and spectral refinement \cite{furuya2024continuous,wang2024latent,rahman2024pretraining,cao2025spectralrefiner}. Most relevant to this work, the U-shaped Neural Operator (UNO) introduces a hierarchical encoder--decoder architecture with skip connections: compressed levels represent large-scale structure while decoder stages restore fine detail \cite{rahman2023uno}. This multiscale representation echoes the scale hierarchy intrinsic to turbulent flows and motivates our choice of backbone.

On the temporal side, most operator architectures treat time as input channels, coordinate embeddings, or short autoregressive windows; none of these choices directly constrains the dynamics during rollout, and the resulting forecasts may become overly dissipative or unstable. Autoregressive prediction of flow fields has been studied with physics-constrained recurrent networks \cite{geneva2020modeling}, and training procedures tailored to multi-step rollout have been explored more recently \cite{bengio2015scheduled,lippe2023pde,takamoto2022pdebench}. Recurrent neural operators (RNOs) couple a Fourier operator with recurrent hidden-state updates and improve temporal coherence, though at a high sequential-computation cost \cite{liu2026tipping}. A parallel line with a long tradition in fluid mechanics is the Koopman operator: nonlinear dynamics can be represented as the linear evolution of observables in an infinite-dimensional function space \cite{koopman1931,mezic2005spectral}. This idea has matured into practical tools such as dynamic mode decomposition \cite{schmid2010dynamic,williams2015data} and has become a standard perspective in the analysis of fluid flows \cite{mezic2013analysis}. Learned finite-dimensional Koopman models \cite{lusch2018deep,takeishi2017learning,korda2018linear} and, most relevant to this work, the Koopman Neural Operator (KNO) \cite{xiong2024koopman} bring this principle to PDE solution fields. However, in our reproduced experiments under a unified protocol, the accuracy of \NEW{KNO} remains limited: on the Kolmogorov-flow benchmarks it does not surpass even the \NEW{FNO} baseline.

These observations suggest that the Koopman perspective is attractive, but how to couple it effectively with neural operators remains an open problem. Reflecting on KNO, we identify a structural gap: the Koopman assumption is formulated in an infinite-dimensional observable space, whereas practical computation is always performed on finite grids with finite latent representations. Truncation inevitably leaves a state-dependent residual, and under autoregressive rollout this residual is repeatedly fed back into the model and amplified. We therefore argue that the residual should be compensated explicitly, rather than hoping that a finite-dimensional linear model absorbs it implicitly.

We propose the Compensated Koopman U-shaped Neural Operator (CoKo-UNO) for stable autoregressive prediction of unsteady flows. CoKo-UNO combines a U-shaped spectral backbone for multiscale spatial representation with Koopman-dominated latent propagation. Analyzing the truncation, we find that the required compensation system itself takes the form of an input-dependent state-space update, for which the selective state-space model (SSM) \cite{kalman1960new,gu2022s4,gu2023mamba,zheng2024aliasfree,song2026adaptive} is a natural learnable parameterization; Mamba thus enters the architecture from the analysis, rather than as an ad hoc module stitched onto the backbone. Resolution-adaptive compensatory skip connections align historical encoder features with future decoding, and an overlapping-warmup strategy provides continuous context at rollout-window boundaries. \NEW{Across four benchmark problems, CoKo-UNO achieves the lowest mean rollout error, reducing the error by $85.71\%$ relative to KNO2d and by up to $76.76\%$ relative to the strongest baseline, while requiring only about $41.40\%$ of RNO's training time.}
The main contributions of this work are summarized as follows:
\begin{itemize}
    \item We identify the state-dependent residual of finite-dimensional Koopman truncation as a key error source in autoregressive operator prediction, formalize how it accumulates during rollout, and show that the required compensation dynamics take the form of an input-dependent state-space update, so that a selective SSM is a principled parameterization of the truncation residual.
    \item We instantiate this principle in CoKo-UNO, which combines a multiscale spectral backbone with latent Koopman--SSM compensation, resolution-adaptive compensatory skip connections, and an overlapping-warmup rollout strategy.
    \item We validate the method on four benchmark problems in terms of accuracy, efficiency, and physical diagnostics, and analyze how the gains relate to the error structure of each flow regime.
\end{itemize}

The remainder of this paper is organized as follows. Section~\ref{sec:method} presents the problem formulation and the CoKo-UNO method. Section~\ref{sec:experiments} reports experiments on Kolmogorov flow, flow past a cylinder, and the shallow-water equations. Section~\ref{sec:conclusion} concludes the paper. Additional flow-field visualizations comparing CoKo-UNO with the baseline models are provided in the appendix.
\section{Method}\label{sec:method}
\subsection{Problem formulation}
We consider a time-dependent partial differential equation of the general form
\begin{equation}
    \partial_t u(\boldsymbol{x},t) = \mathcal{A}(u)(\boldsymbol{x},t),
    \qquad \boldsymbol{x}\in\Omega,\quad t\in(0,T],
    \label{eq:pde}
\end{equation}
\NEW{Here, }$\mathcal{A}$\NEW{ is a possibly strongly nonlinear spatial differential operator. Together with appropriate initial and boundary conditions, the well-posed problem associated with Eq.~}\eqref{eq:pde}\NEW{ induces a solution operator }\NEWMATH{\mathcal{S}_t}\NEW{ that maps the initial field }\NEWMATH{u(\cdot,0)}\NEW{ to the corresponding field }\NEWMATH{u(\cdot,t)}\NEW{ at any later time }\NEWMATH{t\in(0,T]}.
\begin{equation}
    \mathcal{S}_t: u(\cdot,0) \mapsto u(\cdot,t).
    \label{eq:solution_operator}
\end{equation}
In the deep-learning setting, we work with spatially and temporally discretized snapshots of this evolution, produced by a numerical solver, and aim to learn a surrogate that advances the discretized state in time.

\NEW{Let }$u_t\in\mathbb{R}^{C\times H\times W}$\NEW{ denote the discretized system state at time step }$t$\NEW{, where }$C$\NEW{ is the number of physical channels and }$H\times W$\NEW{ is the spatial grid. All methods use an overlapping-warmup autoregressive protocol. Given the most recent }$T_{\mathrm{in}}$\NEW{ states, we form the input window}
\begin{equation}
X_t=[u_{t-T_{\mathrm{in}}+1},\ldots,u_t]\in\mathbb{R}^{T_{\mathrm{in}}\times C\times H\times W}.
\label{eq:history}
\end{equation}
\NEW{Let }$k$\NEW{ be the warmup length and let }$s=T_{\mathrm{in}}-k$\NEW{ be the rollout stride. Each forward pass produces }$\widehat{Y}_t$\NEW{ and is supervised by an aligned }$T_{\mathrm{in}}$\NEW{-frame target block }$Y_t$\NEW{,}
\begin{equation}
Y_t=[u_{t-k+1},\ldots,u_t, u_{t+1},\ldots,u_{t+s}]
\in\mathbb{R}^{T_{\mathrm{in}}\times C\times H\times W}.
\label{eq:target}
\end{equation}
\NEW{The first }$k$\NEW{ entries of }$Y_t$\NEW{ overlap with known history and are used only to warm up the temporal model; only the final }$s$\NEW{ entries, }$[u_{t+1},\ldots,u_{t+s}]$\NEW{, are retained as new predictions and included in the loss. Thus, }$T_{\mathrm{out}}$\NEW{ denotes the total evaluation horizon rather than the output length of one forward pass, and the rollout uses }$N_{\mathrm{AR}}=\lceil T_{\mathrm{out}}/s\rceil$\NEW{ autoregressive windows. Consequently, the learned operator must capture both the spatial structure of the field and its temporal evolution while limiting the accumulation of errors caused by recursive feedback.}

\subsection{Finite-dimensional Koopman truncation}
\label{sec:koopman_truncation}
Consider the discrete-time dynamical system induced by Eq.~\eqref{eq:pde},
\begin{equation}
    x_{t+1}=F(x_t),
    \label{eq:dynamical_system}
\end{equation}
where $x_t$ denotes the system state and $F$ the (generally nonlinear) flow map over one time step. Koopman theory considers, instead of the state itself, scalar observables $g$ of the state and propagates them with the Koopman operator $\mathcal{K}$,
\begin{equation}
    (\mathcal{K}g)(x)=g(F(x)).
    \label{eq:koopman_operator}
\end{equation}
The operator $\mathcal{K}$ is linear by construction, but it acts on an infinite-dimensional space of observables \cite{koopman1931,mezic2005spectral}. This is the key assumption of the Koopman framework: exact linear propagation is only guaranteed in an infinite-dimensional observable space.

Practical computation, however, is always performed on finite grids with a finite-dimensional observable vector $\phi(x)\in\mathbb{R}^{N}$, for example one produced by a learned encoder. A finite-dimensional propagation matrix $K_N\in\mathbb{R}^{N\times N}$ can then only approximate the action of $\mathcal{K}$, and the approximation is exact only if the chosen observables span a Koopman-invariant subspace. In general, the evolution of the truncated representation obeys
\begin{equation}
    \phi(x_{t+1}) = K_N\phi(x_t)+r(x_t),
    \label{eq:koopman_residual}
\end{equation}
where $K_N$ is the truncated Koopman matrix, $K_N\phi(x_t)$ is the finite-dimensional linear propagation term, and $r(x_t)$ is the state-dependent truncation residual. The residual collects the contribution of truncated observable modes and unresolved nonlinear interactions; it is not fixed noise but changes dynamically as the state evolves.

The role of the residual becomes clear when one examines autoregressive rollout. \NEW{Let }$\widehat{\phi}_t$\NEW{ denote the latent state at time }$t$\NEW{ obtained by a model that propagates the state using only the truncated linear operator}, i.e., $\widehat{\phi}_{t+1}=K_N\widehat{\phi}_t$, and define the latent error $e_t=\phi(x_t)-\widehat{\phi}_t$. \NEW{To analyze how this error propagates through autoregressive prediction, we compare the one-step evolutions of the true and predicted latent states.} Subtracting the model recursion from Eq.~\eqref{eq:koopman_residual} gives the error recursion $e_{t+1}=K_N e_t+r(x_t)$, so that
\begin{equation}
    e_T = K_N^{T} e_0 + \sum_{j=0}^{T-1} K_N^{\,T-1-j}\, r(x_j).
    \label{eq:error_recursion}
\end{equation}
Equation~\eqref{eq:error_recursion} shows explicitly how latent errors accumulate along the autoregressive trajectory, and makes two points precise. First, in one-step prediction the residual appears only as a local approximation error. Second, during rollout the residual is injected at every step and propagated through all subsequent states: even when the spectrum of $K_N$ is constrained to the unit disk, so that the linear term does not amplify the error, the accumulated residual contributions do not decay. Stability constraints on the propagation operator, which are commonly imposed in finite-dimensional Koopman models to prevent specific modes from being continuously amplified, therefore control only the linear part of Eq.~\eqref{eq:error_recursion} and leave the residual injection untouched. Moreover, since each predicted state is fed back as input, the trajectory gradually drifts away from the training distribution, which tends to enlarge $r(x_t)$ itself. This observation motivates the design of this work: rather than constraining the linear propagation alone, we explicitly compensate the state-dependent residual $r(x_t)$ while preserving the dominant dynamical structure carried by $K_N$.

\subsection{Selective State-Space Modeling}
\label{sec:ssm_compensation}
State-space models (SSMs) provide a classical framework for modeling temporal dependence \cite{kalman1960new}. A continuous-time linear SSM evolves a latent state $m(t)$ under an input signal $x(t)$ according to
\begin{equation}
    m'(t)=A\,m(t)+B\,x(t),
    \qquad
    y(t)=C\,m(t)+D\,x(t),
    \label{eq:ssm_continuous}
\end{equation}
with state, input, and output matrices $A$, $B$, $C$, and a feedthrough term $D$. Discretizing Eq.~\eqref{eq:ssm_continuous} with a timescale parameter $\Delta$ under the zero-order-hold rule gives a discrete-time update $m_k=\overline{A}m_{k-1}+\overline{B}x_k$ with $\overline{A}=\exp(\Delta A)$ and $\overline{B}=(\Delta A)^{-1}(\exp(\Delta A)-I)\,\Delta B$, which structured SSMs parameterize and compute efficiently \cite{gu2022s4}.

Selective SSMs go one step further and make the discretization and input parameters depend on the current input \cite{gu2023mamba}, i.e., $\Delta=\Delta(x_k)$, $B=B(x_k)$, and $C=C(x_k)$:
\begin{equation}
    m_k=\overline{A}(x_k)\,m_{k-1}+\overline{B}(x_k)\,x_k,
    \qquad
    y_k=\overline{C}(x_k)\,m_k+\overline{D}\,x_k.
    \label{eq:selective_ssm}
\end{equation}
\NEW{Unlike conventional SSMs with fixed transition dynamics, a selective SSM adapts its state update to the current input. In Eq.~}\eqref{eq:selective_ssm}\NEW{, }$\overline{A}(x_k)$\NEW{ adjusts the state transition, while }$\overline{B}(x_k)x_k$\NEW{ injects state-dependent information into the latent state. This input-dependent update provides the residual contribution that is absent from fixed finite-dimensional Koopman propagation. CoKo-UNO implements the selective SSM using Mamba and employs it to compensate the truncation residual described in Section~}\ref{sec:koopman_truncation}\NEW{; its resolution-aware deployment is specified in Section~2.4.}
\subsection{CoKo-UNO architecture}
\label{sec:architecture}
\begin{figure}[H]
\centering
\makebox[0pt][c]{\includegraphics[width=\paperwidth,trim=0 512 0 0,clip]{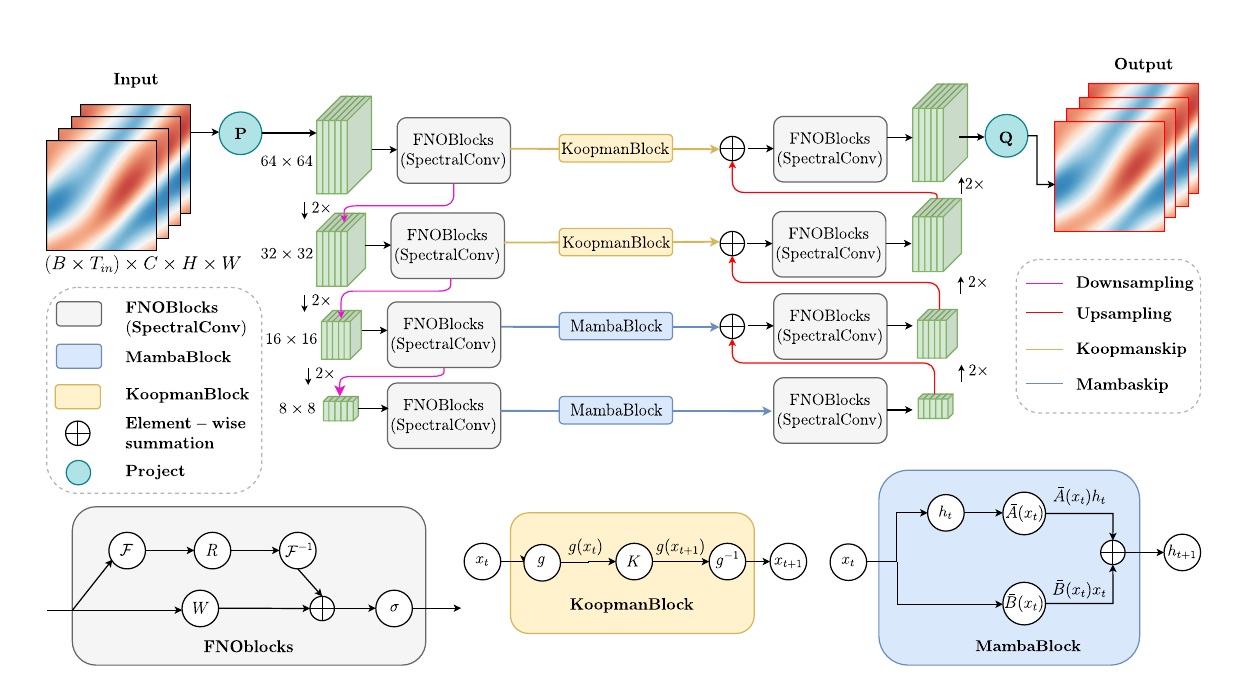}}
\caption{\NEW{Architecture of the Compensated Koopman U-shaped Neural Operator (CoKo-UNO). The U-shaped spectral backbone extracts and reconstructs multiscale spatial features. Mamba performs selective state-space compensation at the bottleneck and low-resolution skip levels, whereas Koopman modules apply efficient linear processing at high-resolution skip levels before decoder fusion.}}
\label{fig:architecture}
\end{figure}

\begin{figure}[H]
\centering
\makebox[0pt][c]{\includegraphics[width=\paperwidth,trim=0 725 0 25,clip]{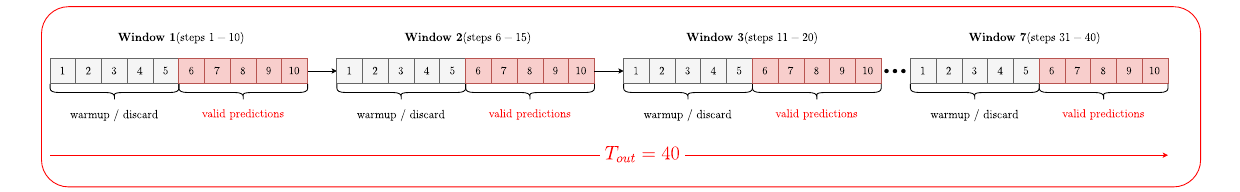}}
\caption{\NEW{Overlapping autoregressive warmup strategy, illustrated for Kolmogorov flow with $\nu=10^{-3}$ ($T_{\mathrm{in}}=10$, $T_{\mathrm{out}}=40$). Adjacent rollout windows retain overlapping historical states for warmup, and only the newly generated states are used as valid predictions.}}
\label{fig:overlapping_warmup}
\end{figure}

\NEW{We now introduce the CoKo-UNO architecture. Given the history window }$X_t$\NEW{, a U-shaped spectral backbone first encodes the input into multiscale features. In CoKo-UNO, Mamba implements the selective-SSM compensation mechanism described in Section~2.3. It is deployed at the bottleneck and selected low-resolution skip levels, whereas high-resolution skip features use Koopman-type linear compensation for efficient feature processing. The processed multiscale features are subsequently fused by the decoder to reconstruct the future fields }$\widehat{Y}_t$\NEW{. Figs.~}\ref{fig:architecture}\NEW{ and }\ref{fig:overlapping_warmup}\NEW{ provide overviews of the architecture and rollout strategy, respectively.}

\vspace{-24pt}

\subsubsection{U-shaped multiscale spatial representation based on spectral operators}
The spatial backbone of CoKo-UNO adopts a U-shaped neural-operator architecture consisting of an encoder, a bottleneck, and a decoder. Let $z^{(l)} \in \mathbb{R}^{T_{\mathrm{in}}\times C_l\times H_l\times W_l}$ denote the feature representation at level $l$, where $l=0,\ldots,L$. The initial feature is defined as $z^{(0)} = X_{t-T_{\mathrm{in}}+1:t}$. The encoder recursively applies spectral operator blocks according to
\begin{equation}
    z^{(l+1)} = \mathcal{F}^{\mathrm{enc}}_l \left( z^{(l)} \right),
    \qquad l=0,\ldots,L-1,
\end{equation}
where $\mathcal{F}^{\mathrm{enc}}_l$ denotes a spectral block with target spatial resolution $H_{l+1}\times W_{l+1}$. For an input feature $u\in\mathbb{R}^{C\times H\times W}$, the spectral convolution operator is defined as
\begin{equation}
    \mathcal{K}_l(u) = \mathcal{F}^{-1} \left( R_l\,\mathcal{F}(u) \right),
\end{equation}
where $\mathcal{F}$ and $\mathcal{F}^{-1}$ denote the Fourier transform and inverse Fourier transform, respectively, and $R_l$ is a learnable complex-valued weight defined on a set of retained Fourier modes. Specifically, if $\widehat{u}(\xi)$ is the Fourier coefficient of $u$ at frequency $\xi$, then
\begin{equation}
    \widehat{v}(\xi)
    =
    \begin{cases}
        R_l(\xi)\widehat{u}(\xi), & \xi\in\Omega_l, \\
        0, & \xi\notin\Omega_l,
    \end{cases}
\end{equation}
where $\Omega_l$ denotes the set of retained frequency modes. Accordingly, the $l$th spectral block is written as
\begin{equation}
    \mathcal{F}^{\mathrm{enc}}_l(u) = \sigma \left( \mathcal{K}_l(u)+W_lu \right),
\end{equation}
where $W_l$ is a pointwise linear transformation and $\sigma(\cdot)$ is a nonlinear activation function. The decoder restores the spatial resolution in the reverse direction. Let $s^{(l)}$ denote the skip feature at level $l$. The decoder features are updated as
\begin{equation}
    \NEWMATH{z^{(l)} = \mathcal{F}^{\mathrm{dec}}_l \left( \left[z^{(l+1)}\,\Vert\,s^{(l)}\right] \right),}
    \qquad l=L-1,\ldots,0.
\end{equation}
\NEW{Finally, a projection layer }$\mathcal{Q}$\NEW{ maps the decoded representation to the aligned output block:}
\begin{equation}
    \NEWMATH{\widehat{Y}_{t} = \mathcal{Q} \left( z^{(0)} \right).}
\end{equation}

\NEW{The preceding formulation provides multiscale spatial encoding and reconstruction. The temporal compensation mechanism is introduced subsequently: Mamba processes the bottleneck and selected low-resolution skip features, whereas high-resolution skip features undergo Koopman-type linear processing.}

\subsubsection{Latent Koopman--SSM Compensation}

\NEW{Finite-dimensional Koopman propagation retains the dominant linear component of the dynamics but leaves the state-dependent truncation residual identified in Section~}\ref{sec:koopman_truncation}\NEW{ unresolved. The selective SSM of Eq.~}\eqref{eq:selective_ssm}\NEW{ provides the missing input-dependent residual-update structure through the term }$\overline{B}(x_k)x_k$\NEW{. Mamba implements this selective SSM and is therefore used as the learnable compensation mechanism in CoKo-UNO.}

\NEW{To balance compensation capability and computational cost across resolutions, Mamba-based selective-SSM compensation is deployed at the bottleneck and low-resolution skip levels. High-resolution skip features instead use Koopman-type linear compensation to process detailed spatial features efficiently.}

\NEW{Let }$z^{(b)}\equiv z^{(L)}\in\mathbb{R}^{T_{\mathrm{in}}\times C_b\times H_b\times W_b}$\NEW{ denote the bottleneck feature produced by the UNO encoder. Since it has the coarsest spatial resolution, }$z^{(b)}$\NEW{ provides a compact representation of the global dynamics in the input history. The bottleneck is processed by Mamba as}
\begin{equation}
    \NEWMATH{z_{\mathrm{M}}^{(b)}
    =
    \operatorname{Restore}\!\left[
    \operatorname{Mamba}\!\left(
    \operatorname{Seq}\!\left(z^{(b)}\right)
    \right)
    \right].}
    \label{eq:bottleneck_mamba}
\end{equation}
\NEW{Here, }$\operatorname{Seq}(\cdot)$\NEW{ rearranges the bottleneck feature into temporal sequences of size }$(BH_bW_b)\times T_{\mathrm{in}}\times C_b$\NEW{. Mamba then applies the selective state update of Eq.~}\eqref{eq:selective_ssm}\NEW{ along these sequences. In particular, the input-dependent injection term provides the residual-update structure that is absent from fixed finite-dimensional Koopman propagation. }$\operatorname{Restore}(\cdot)$\NEW{ maps the processed sequences back to the bottleneck feature layout. The resulting feature }$z_{\mathrm{M}}^{(b)}$\NEW{ is used as the selectively compensated global representation for decoding.}

\NEW{For the skip pathways, the }$64\times64$\NEW{ main experiments use partial compensation. Let }$\rho_l=\min(H_l,W_l)$\NEW{ denote the minimum spatial dimension of the encoder feature }$z^{(l)}$\NEW{, and let }$\tau$\NEW{ be a resolution threshold. The processing module for a skip feature is selected as}
\begin{equation}
    \NEWMATH{\mathcal{T}_l\!\left(z^{(l)}\right)=
    \begin{cases}
    \mathcal{M}_l\!\left(z^{(l)}\right), & \rho_l\leq\tau,\\
    \mathcal{K}_l\!\left(z^{(l)}\right), & \rho_l>\tau,
    \end{cases}}
    \label{eq:module_allocation}
\end{equation}
\NEW{where }$\mathcal{M}_l$\NEW{ and }$\mathcal{K}_l$\NEW{ denote the Mamba-based selective-SSM compensation module and the Koopman-type linear compensation module, respectively. Thus, low-resolution skip features receive the input-dependent residual-update structure of Mamba, whereas high-resolution skip features are processed by Koopman-type linear compensation.}

\NEW{Although selective-SSM compensation can in principle be applied to every skip pathway, deploying Mamba at high spatial resolutions is computationally expensive because the corresponding feature maps must be processed as long sequences. We therefore retain Mamba at the bottleneck and low-resolution skip levels, where the compact feature maps make selective state-space modeling affordable, and use Koopman-type linear compensation at high-resolution levels to preserve efficient processing of fine-scale spatial features.}

\NEW{Equation~}\eqref{eq:module_allocation}\NEW{ specifies the partial-compensation configuration used in the main }$64\times64$\NEW{ experiments. Full compensation and all-linear compensation are evaluated in the ablation study at }$32\times32$\NEW{.}

\subsubsection{\NEW{Resolution-adaptive skip connections}}
\label{sec:skip}
\NEW{Encoder features contain multiscale spatial information. In CoKo-UNO, the retained encoder features are not directly concatenated with decoder features. Instead, each skip feature is first processed by the resolution-aware module before decoder fusion. For the encoder feature }$z^{(l)}$\NEW{ at level }$l$\NEW{, the processed skip feature is defined as}
\begin{equation}
\NEWMATH{s^{(l)}=\operatorname{Align}_l\!\left(\mathcal{T}_l\!\left(z^{(l)}\right)\right).}
\label{eq:skip_alignment}
\end{equation}
\NEW{Here, }$\mathcal{T}_l$\NEW{ denotes the Mamba or Koopman module selected according to the spatial resolution of the source feature map. The operator }$\operatorname{Align}_l$\NEW{ resamples the processed skip feature to match the spatial resolution of the corresponding decoder stage. For low-resolution skip features, the Mamba module performs selective state-space modeling along the temporal dimension. For high-resolution skip features, the Koopman module applies a shared learnable linear transformation to the channel features at each time step and spatial location. The aligned feature }$s^{(l)}$\NEW{ is then concatenated with the current decoder feature along the channel dimension and processed by the corresponding UNO decoder block. Thus, each decoder stage fuses spatially aligned Mamba-processed low-resolution features, which carry the input-dependent residual-compensation structure, with Koopman-processed high-resolution features that preserve detailed spatial information through efficient linear compensation.}

\begin{table}[H]
\centering
\caption{\NEW{Selective-SSM and Koopman-type linear compensation strategies for the bottleneck and skip pathways. Partial compensation is used in the $64\times64$ main experiments, whereas full and all-linear compensation are evaluated in the ablation study at $32\times32$.}}
\label{tab:compensation_strategies}
\resizebox{\textwidth}{!}{%
\begin{tabular}{lcc}
\toprule
\NEW{Strategy} & \NEW{Selective-SSM compensation} & \NEW{Koopman-type linear compensation} \\
\midrule
\NEW{Full compensation}     & \NEW{bottleneck + all skip levels} & --- \\
\NEW{Partial compensation}  & \NEW{bottleneck + low-resolution skip levels} & \NEW{high-resolution skip levels} \\
\NEW{All-linear compensation} & --- & \NEW{bottleneck + all skip levels} \\

\bottomrule
\end{tabular}}
\end{table}
\subsubsection{Overlapping-warmup autoregressive rollout}
\NEW{Long forecasts are generated by the overlapping-warmup protocol defined above. Each forward pass takes }$T_{\mathrm{in}}$\NEW{ frames as input and returns an aligned }$T_{\mathrm{in}}$\NEW{-frame output block. Its first }$k$\NEW{ frames overlap with already available history and are used only to warm up the temporal model, while its final }$s=T_{\mathrm{in}}-k$\NEW{ frames are accepted as new predictions. The input window then advances by }$s$\NEW{ frames. Consequently, adjacent windows share }$k$\NEW{ context frames, and the total rollout horizon }$T_{\mathrm{out}}$\NEW{ is obtained after }$\lceil T_{\mathrm{out}}/s\rceil$\NEW{ windows. This protocol is applied identically to CoKo-UNO and all baselines. The overlapping frames provide continuous temporal context---in particular for the hidden state of the selective SSM, which would otherwise be re-initialized from self-generated predictions at every window boundary---and are excluded from both the loss and the reported prediction error.}
\subsection{Training objective and protocol}
\label{sec:training}
The model is trained on snapshot sequences produced by numerical solvers. Training samples are windows $(X_t,Y_t)$ as defined in Eqs.~\eqref{eq:history} and \eqref{eq:target}, sampled from the training trajectories. \NEW{The model is trained by direct multi-step autoregressive prediction without teacher forcing. Let }$N_{\mathrm{AR}}=\lceil T_{\mathrm{out}}/s\rceil$\NEW{ denote the number of rollout windows, where }$s=T_{\mathrm{in}}-k$\NEW{ is the number of newly predicted frames retained from each window and }$k$\NEW{ is the warmup (overlap) length. For the }$r$\NEW{th rollout window, let }$B$\NEW{ denote the mini-batch size and let }$q=1,\ldots,s$\NEW{ index the newly predicted frames that participate in supervision. The prediction loss is defined as}
\begin{equation}
    \NEWMATH{\mathcal{L}_{\mathrm{pred}}
    =
    \sum_{r=1}^{N_{\mathrm{AR}}}
    \sum_{i=1}^{B}
    \sum_{q=1}^{s}
    \frac{\left\|\widehat{u}_{i,q}^{(r)}-u_{i,q}^{(r)}\right\|_{F}}
         {\left\|u_{i,q}^{(r)}\right\|_{F}}},
    \label{eq:prediction_loss}
\end{equation}
\NEW{Here, }$\widehat{u}_{i,q}^{(r)}$\NEW{ and }$u_{i,q}^{(r)}$\NEW{ denote the predicted and reference fields, respectively, for the }$q$\NEW{th valid output of the }$r$\NEW{th rollout window and the }$i$\NEW{th sample in the mini-batch. The Frobenius norm is evaluated over all spatial locations and physical channels of an individual field. The first }$k$\NEW{ outputs of each window are used only for warmup and are excluded from the prediction loss.}

\NEW{For training monitoring, we report the epoch-averaged training relative }$L_2$\NEW{ error, obtained by averaging the per-frame relative }$L_2$\NEW{ errors over all training samples and all }$T_{\mathrm{out}}$\NEW{ valid prediction steps. This quantity measures the mean training prediction error over an epoch, whereas }$\mathcal{L}_{\mathrm{pred}}$\NEW{ in Eq.~}\eqref{eq:prediction_loss}\NEW{ is the prediction loss used for backpropagation.}

\NEW{To additionally supervise the multiscale latent features, let }$\mathcal{H}^{(r)}$\NEW{ denote the set of selected features in the }$r$\NEW{th rollout window, including the bottleneck and the selected skip features. The reconstruction loss is}
\begin{equation}
    \NEWMATH{\mathcal{L}_{\mathrm{rec}}^{(r)}
    =
    \frac{1}{\left\lvert\mathcal{H}^{(r)}\right\rvert}
    \sum_{h_i^{(r)}\in\mathcal{H}^{(r)}}
    \left\|\widehat{h}_i^{(r)}-h_i^{(r)}\right\|_2^2},
    \label{eq:single_reconstruction_loss}
\end{equation}
\NEW{The final training objective combines the prediction loss in Eq.~}\eqref{eq:prediction_loss}\NEW{ and the reconstruction loss in Eq.~}\eqref{eq:single_reconstruction_loss}\NEW{ as}
\begin{equation}
    \NEWMATH{\mathcal{L}
    =
    \lambda_{\mathrm{pred}}\mathcal{L}_{\mathrm{pred}}
    +
    \sum_{r=1}^{N_{\mathrm{AR}}}\lambda_{\mathrm{rec}}^{(r)}\mathcal{L}_{\mathrm{rec}}^{(r)}},
    \label{eq:total_loss}
\end{equation}
\NEW{where }$\lambda_{\mathrm{pred}}$\NEW{ controls the prediction term and }$\lambda_{\mathrm{rec}}^{(r)}$\NEW{ controls the reconstruction contribution of the }$r$\NEW{th rollout window; their numerical values are specified in the experimental settings. Optimization uses Adam with an initial learning rate of $10^{-3}$ and a weight decay of $10^{-4}$ for $200$ epochs; further experimental details are given in}~\ref{app:implementation}\NEW{.}
\section{Experiments}\label{sec:experiments}
\subsection{Experimental setup}
\label{sec:setup}
We evaluate the proposed method on four representative unsteady-flow benchmarks covering vortical flows, bluff-body wake dynamics, and shallow-water wave evolution.

\textbf{Kolmogorov flow.} Two-dimensional incompressible Navier--Stokes flow in vorticity form on the periodic unit square, driven by the Kolmogorov forcing $f(\boldsymbol{x})=0.1\bigl(\sin(2\pi(x_1+x_2))+\cos(2\pi(x_1+x_2))\bigr)$ \cite{chandler2013invariant}, following the standard neural-operator benchmark setting \cite{li2021fourier,xiong2024koopman}. Two viscosity coefficients, $\nu=10^{-3}$ and $\nu=10^{-4}$, assess the model under increasingly advection-dominated regimes. \NEW{The data are obtained from the official NeuralOperator library, generated at a $256\times256$ resolution, and downsampled to $64\times64$ for our experiments. The physical-time horizons are $T=50$ and $T=30$, respectively. Detailed data-generation settings are provided in Appendix}~\ref{app:implementation}\NEW{.}

\NEW{\textbf{Cylinder flow.} Two-dimensional incompressible flow past a circular cylinder from the CFDBench benchmark suite} \cite{luo2023cfdbench}\NEW{, evaluated using the velocity magnitude $\lvert\mathbf{U}\rvert$. A uniform horizontal velocity inlet $(u,v)=(U_{\mathrm{in}},0)$ is imposed at the left boundary, the upper and lower walls and the cylinder surface satisfy no-slip conditions, and the right boundary is an outlet. The case-specific metadata specify $U_{\mathrm{in}}$, density $\rho$, dynamic viscosity $\mu$, the rectangular-domain bounds, and cylinder radius; the transient fields are generated with an internal and exported time step of $\Delta t=10^{-3}\,\mathrm{s}$ and interpolated to a $64\times64$ Cartesian grid. This case focuses on the dynamically sensitive wake and vortex-shedding region downstream of the body} \cite{williamson1996vortex}\NEW{.}

\textbf{Shallow-water equations (SWE).} A two-dimensional shallow-water benchmark from PDEBench \cite{takamoto2022pdebench}, of the type also used in KNO evaluations \cite{xiong2024koopman}, included to examine the applicability of the method to wave-propagation dynamics. The predicted field is the water depth. \NEW{The data simulate a radial dam-break problem on $[-2.5,2.5]^2$: the initial water depth is $h=2$ within a disk centered at the origin and $h=1$ outside it, with zero initial momentum. The disk radius is sampled independently from $[0.3,0.7]$ for each trajectory. Homogeneous zero-gradient (extrapolation) boundary conditions are imposed on all four boundaries. Trajectories are saved over $T=1.0$ at 100 uniform intervals ($\Delta t=0.01$) and are generated at $128\times128$ resolution.}
For each dataset, $1{,}000$ samples are used for training and $200$ samples for testing. \NEW{All experiments use $T_{\mathrm{in}}=10$ input time steps and autoregressively predict the subsequent $T_{\mathrm{out}}$ steps: $T_{\mathrm{out}}=40$ for Kolmogorov flow with $\nu=10^{-3}$, $T_{\mathrm{out}}=20$ for Kolmogorov flow with $\nu=10^{-4}$, and $T_{\mathrm{out}}=10$ for both the cylinder-flow and SWE datasets.} The main comparative experiments are conducted at a spatial resolution of $64\times64$, with additional experiments at $32\times32$ for the ablation study \NEW{described in Section~}\ref{sec:ablation}\NEW{.} All models are trained for $200$ epochs using Adam with an initial learning rate of $10^{-3}$ and a weight decay of $10^{-4}$, and all methods are evaluated under the same overlapping-window autoregressive protocol, in which consecutive prediction windows share five time steps and the rollout advances with a stride of five. \NEW{For every dataset and model, we conduct three independent training runs with random seeds $42$, $2025$, and $3407$. Unless stated otherwise, quantitative accuracy results are reported as the mean $\pm$ standard deviation of the test-set mean full-rollout errors over these three runs.}

\NEW{Performance is evaluated using the mean relative $L_2$ error of the full rollout over the test set,}
\begin{equation}
    \NEWMATH{\bar{e}
    =
    \frac{1}{N_{\mathrm{test}}}\sum_{i=1}^{N_{\mathrm{test}}}
    \frac{\left\|\widehat{\mathbf{u}}^{(i)}-\mathbf{u}^{(i)}\right\|_2}
         {\left\|\mathbf{u}^{(i)}\right\|_2},}
    \label{eq:mean_rel_l2}
\end{equation}
\NEW{Here, }$\widehat{\mathbf{u}}^{(i)}$\NEW{ and }$\mathbf{u}^{(i)}$\NEW{ denote the predicted and reference trajectories obtained by concatenating all frames of the }$i$\NEW{th test rollout. The norm is therefore evaluated jointly over the temporal trajectory, spatial locations, and physical channels. The relative }$L_2$\NEW{ metric is standard in neural-operator benchmarks}~\cite{li2021fourier,takamoto2022pdebench}\NEW{; its step-wise counterpart is used in Section~}\ref{sec:longhorizon}\NEW{ to resolve error accumulation over time.}

Four baselines are compared. \NEW{In the following, FNO and KNO refer to the original operator formulations, whereas FNO2d and KNO2d denote their respective two-dimensional baseline implementations used in our experiments.} FNO2d \cite{li2021fourier}, KNO2d \cite{xiong2024koopman}, and RNO \cite{liu2026tipping} are representative Fourier-based, Koopman-based, and recurrent neural operators, respectively. The standard UNO \cite{rahman2023uno}, which concatenates historical states along the channel dimension, \NEW{is denoted UNO T-channel and} identifies how much of the performance is attributable to the multiscale backbone alone. An additional UNO T-batch control, which folds the temporal dimension into the batch dimension and applies a shared UNO independently to each time step, is considered in the ablation study \NEW{described in Section~}\ref{sec:ablation}\NEW{.} All baselines are retrained under the identical training and rollout protocol. Implementation details, the hardware environment, and data and code availability are documented in Appendix~\ref{app:implementation}. \NEW{Table~}\ref{tab:complexity}\NEW{ summarizes the model complexity on the Kolmogorov flow dataset with $\nu = 10^{-3}$. Compared with FNO2d, CoKo-UNO reduces the peak memory consumption by $43.47\%$. Although its peak memory usage is higher than that of RNO, CoKo-UNO reduces the per-step inference time by $64.76\%$, indicating a substantially more efficient inference process.}

\begin{table}[H]
\centering
\small
\setlength{\tabcolsep}{4pt}
\caption{Model complexity comparison.}
\label{tab:complexity}
\resizebox{\textwidth}{!}{%
\begin{tabular}{lccc}
\toprule
Method & Parameters (M) & Peak memory (MiB) & Inference time per step (ms) \\
\midrule
FNO2d          & 1.19 & $1032.77 \pm 0.00$ & $1.27 \pm 0.01$ \\
KNO2d          & 0.15 & $914.35 \pm 0.00$  & $1.23 \pm 0.00$ \\
RNO            & 2.12 & $440.73 \pm 0.00$  & $32.75 \pm 0.14$ \\
UNO T-channel  & 4.30 & $1048.53 \pm 0.42$ & $12.01 \pm 0.06$ \\
UNO T-batch    & 4.30 & $659.32 \pm 0.55$  & $11.99 \pm 0.05$ \\
CoKo-UNO       & 4.64 & $583.86 \pm 4.67$  & $11.54 \pm 0.05$ \\
\bottomrule
\end{tabular}%
}
\end{table}

\subsection{Benchmark results}
\label{sec:benchmark}
\begin{table}[!htbp]
\centering
\caption{Mean relative $L_2$ rollout errors on the four benchmark problems at $64\times64$ resolution. \NEW{Entries are the mean $\pm$ standard deviation of the test-set mean full-rollout errors over three independent runs (random seeds $42$, $2025$, and $3407$). Lower values indicate better performance; the best mean in each column is in bold.}}
\label{tab:main_results}
\begingroup
\fontsize{10}{12}\selectfont
\setlength{\tabcolsep}{0pt}
\renewcommand{\arraystretch}{1.12}
\makebox[\textwidth][c]{%
\begin{tabular}{>{\centering\arraybackslash}m{2.5cm}>{\centering\arraybackslash}m{3.9cm}>{\centering\arraybackslash}m{3.9cm}>{\centering\arraybackslash}m{3.9cm}>{\centering\arraybackslash}m{3.9cm}}
\toprule
Method & Kolmogorov $\nu=10^{-3}$ & Kolmogorov $\nu=10^{-4}$ & Cylinder $\lvert\mathbf{U}\rvert$ & SWE \\
\midrule
FNO2d         & \meanstd{4.76}{-2}{9.97}{-4} & \meanstd{2.57}{-1}{1.60}{-3} & \meanstd{1.94}{-2}{1.50}{-3} & \meanstd{1.35}{-2}{2.28}{-3} \\
KNO2d         & \meanstd{7.63}{-2}{2.45}{-3} & \meanstd{3.20}{-1}{1.54}{-3} & \meanstd{4.50}{-2}{4.65}{-3} & \meanstd{1.33}{-2}{7.90}{-4} \\
RNO           & \meanstd{8.66}{-2}{8.60}{-2} & \meanstd{2.01}{-1}{2.99}{-2} & \meanstd{1.27}{-2}{1.43}{-3} & \meanstd{2.50}{-3}{1.91}{-4} \\
UNO T-channel & \meanstd{4.69}{-2}{1.11}{-3} & \meanstd{2.54}{-1}{2.05}{-3} & \meanstd{1.88}{-2}{5.76}{-4} & \meanstd{5.95}{-3}{1.08}{-3} \\
\textbf{CoKo-UNO}
         & \bestmeanstd{1.09}{-2}{1.03}{-4}
         & \bestmeanstd{1.19}{-1}{1.87}{-3}
         & \bestmeanstd{1.05}{-2}{3.39}{-4}
         & \bestmeanstd{2.01}{-3}{2.93}{-5} \\
\bottomrule
\end{tabular}
}
\endgroup
\end{table}

\NEW{Table~}\ref{tab:main_results}\NEW{ shows that, based on the mean errors, CoKo-UNO achieves the lowest rollout error on all four benchmark problems, establishing it as the best-performing model in this comparison. Three observations are worth discussing in detail.}

\NEW{First, the comparison with the non-recurrent operators isolates the role of explicit temporal modeling. KNO2d is the only Koopman-based baseline in the comparison. Although it incorporates latent linear propagation, its mean rollout errors are higher than those of FNO2d on both Kolmogorov-flow cases. This observation suggests that, under the present autoregressive setting, latent linear propagation may benefit from an additional mechanism for handling the accumulated truncation residuals. On Kolmogorov flow with $\nu=10^{-3}$, CoKo-UNO reduces the mean error from $7.63\times10^{-2}$ to $1.09\times10^{-2}$, an $85.71\%$ reduction relative to KNO2d. This is consistent with the analysis of Section~}\ref{sec:koopman_truncation}\NEW{: a finite-dimensional Koopman model without residual compensation leaves the state-dependent truncation error uncorrected, and under autoregressive rollout this error accumulates according to Eq.~}\eqref{eq:error_recursion}\NEW{. The UNO T-channel baseline confirms that the multiscale backbone alone is likewise insufficient for stable long-term prediction.}

\NEW{Second, the strongest baseline differs across the four problems: UNO T-channel has the lowest mean baseline error on Kolmogorov flow with $\nu=10^{-3}$, whereas RNO has the lowest mean baseline error on the remaining three problems. We therefore compare CoKo-UNO with the strongest baseline for each case. Relative to these baselines, CoKo-UNO reduces the mean rollout error by $76.76\%$ and $40.80\%$ on the two Kolmogorov-flow cases, by $17.32\%$ for the cylinder-flow velocity magnitude $\lvert\mathbf{U}\rvert$, and by $19.60\%$ on SWE.} Since RNO is by far the most expensive model to train, a comparison of computational cost is deferred to the accuracy--time trade-off analysis \NEW{in Section~}\ref{sec:tradeoff}\NEW{, where CoKo-UNO is shown to require only about $41.40\%$ of RNO's training time.}

\NEW{Third, the magnitude of the improvement varies considerably across problems, and this variation is best explained by the error structure of each case rather than by a single dimensionless number. On cylinder flow and SWE, the strongest baselines already attain very small mean rollout errors ($1.27\times10^{-2}$ for the velocity magnitude $\lvert\mathbf{U}\rvert$ and $2.50\times10^{-3}$ for SWE, respectively), so the headroom for further reduction is inherently limited; the mean-error reductions of $17.32\%$ and $19.60\%$ on these problems are nevertheless consistent, and Section~}\ref{sec:diagnostics}\NEW{ shows that they are retained in the physically sensitive regions (the cylinder wake and the conserved water mass).} \NEW{On Kolmogorov flow at $\nu=10^{-3}$, the best baseline error is substantially larger and is dominated by error accumulation over the $40$-step rollout, which is precisely the error source that the compensation mechanism acts on, as described in Eq.~}\eqref{eq:error_recursion}\NEW{; hence the largest mean-error reduction ($76.76\%$). At $\nu=10^{-4}$, all neural operators degrade as the effective Reynolds number increases---a widely observed phenomenon consistent with the Green's-function interpretation of neural operators, whose integral kernels are naturally aligned with diffusion-dominated dynamics}~\cite{kovachki2023neural}\NEW{---and the error is increasingly dominated by one-step representation error rather than by its accumulation. Since the compensation mechanism acts on error growth, its relative benefit ($40.80\%$) is necessarily smaller when the dominant error source is representation bias. Thus, CoKo-UNO is most effective when error accumulation is the bottleneck and complements improved spatial representations.}
\subsection{Rollout error accumulation analysis}
\label{sec:longhorizon}
To examine how prediction errors propagate in time, we evaluate the step-wise relative $L_2$ error
\begin{equation}
    e_t
    =
    \frac{1}{N_{\mathrm{test}}}
    \sum_{i=1}^{N_{\mathrm{test}}}
    \frac{\left\|\hat{u}^{(i)}_t-u^{(i)}_t\right\|_2}
         {\left\|u^{(i)}_t\right\|_2},
    \label{eq:stepwise_error}
\end{equation}
\NEW{Unlike the full-rollout metric in Eq.~}\eqref{eq:mean_rel_l2}\NEW{, this step-wise metric directly characterizes the temporal accumulation of errors during autoregressive rollout.}

\begin{figure}[H]
\centering
\includegraphics[width=0.55\textwidth]{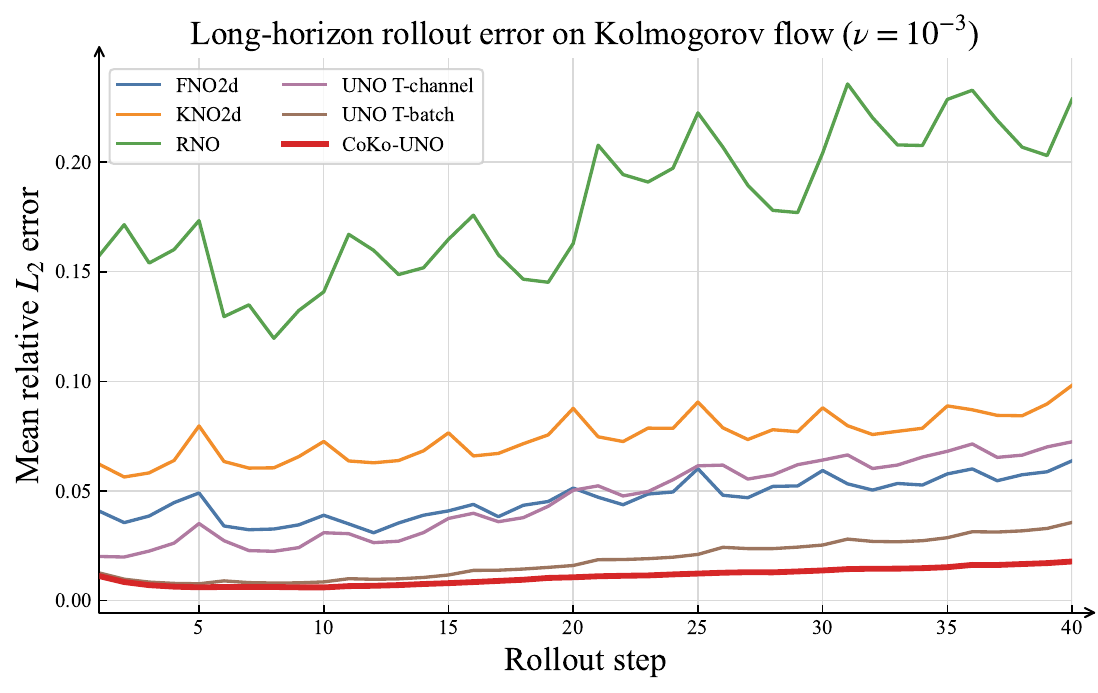}
\caption{Time-resolved mean relative $L_2$ rollout error on Kolmogorov flow with $\nu=10^{-3}$ ($64\times64$). CoKo-UNO maintains both the lowest error level and the slowest error growth over the $40$-step horizon.}
\label{fig:per_step_error}
\end{figure}

\NEW{Figure}~\ref{fig:per_step_error} reports the step-wise error on Kolmogorov flow with $\nu=10^{-3}$, the most demanding rollout in our benchmarks ($40$ steps). \NEW{The curves in Fig.~}\ref{fig:per_step_error}\NEW{ and the fitted parameters in Table~}\ref{tab:growth_rates}\NEW{ are obtained from the runs with random seed $42$ for all methods.} All methods exhibit growing errors as the horizon extends, but the growth rates differ markedly. \NEW{Among the baseline methods, FNO2d has the smallest fitted growth rate; nevertheless, CoKo-UNO achieves a lower fitted growth rate, indicating more effective suppression of error accumulation.} The remaining benchmark problems exhibit the same qualitative behavior and are omitted for brevity.

To quantify error growth rather than error level, we fit the step-wise error curve of each model with the linear law $e_t \approx e_0 + \gamma t$, so that $\gamma$ measures the per-step error-growth rate. The linear law is suggested by Eq.~\eqref{eq:error_recursion}: the propagation operator is residual-parameterized, $K_b = I + \Delta K$ with small $\|\Delta K\|$, and over horizons for which $T\|\Delta K\| \ll 1$ the binomial expansion gives $K_b^{T} \approx I + T\Delta K$; the accumulated error $e_T = K_b^{T}e_0 + \sum_{j} K_b^{\,T-1-j} r(x_j)$ is then dominated by the progressively injected residuals $\sum_j r(x_j)$ and grows approximately linearly in $T$. Two further consequences are worth noting. First, a systematic, state-dependent residual produces a sum that grows linearly in $T$, whereas uncorrelated, noise-like residuals would yield only $\sqrt{T}$ growth; the observed linear growth therefore corroborates the state-dependent nature of the truncation residual assumed in Section~\ref{sec:koopman_truncation}. Second, sustained linear---rather than exponential---growth over the full horizon indicates that no eigenvalue of the learned propagation operator lies significantly outside the unit circle, providing indirect evidence for the near-unit-circle spectrum expected of residual architectures \NEW{discussed in Section~}\ref{sec:architecture}\NEW{.} Table~\ref{tab:growth_rates} reports the fitted growth rates. CoKo-UNO exhibits the smallest $\gamma$ among all compared methods, confirming that its advantage lies primarily in suppressing error accumulation rather than in one-step fitting. This behavior is precisely what the compensated-Koopman interpretation predicts: the selective state-space bottleneck corrects the state-dependent truncation residual at every step, so that the residual injection term in Eq.~\eqref{eq:error_recursion} is reduced throughout the trajectory rather than only near the initial window.

\begin{table}[H]
\centering
\caption{Fitted parameters of the linear error-growth model $e_t\approx e_0+\gamma t$ for step-wise relative $L_2$ error on Kolmogorov flow with $\nu=10^{-3}$. Smaller $\gamma$ indicates slower error accumulation.}
\label{tab:growth_rates}
\begin{tabular}{lcc}
\toprule
Method & Error-growth rate $\gamma$ & Intercept $e_0$ \\
\midrule
FNO2d         & \NEW{$6.57\times10^{-4}$} & \NEW{$3.29\times10^{-2}$} \\
KNO2d         & \NEW{$7.32\times10^{-4}$} & \NEW{$5.95\times10^{-2}$} \\
RNO           & \NEW{$2.26\times10^{-3}$} & \NEW{$1.34\times10^{-1}$} \\
UNO T-channel & \NEW{$1.45\times10^{-3}$} & \NEW{$1.65\times10^{-2}$} \\
UNO T-batch   & \NEW{$7.21\times10^{-4}$} & \NEW{$3.53\times10^{-3}$} \\
CoKo-UNO      & \NEW{$2.96\times10^{-4}$} & \NEW{$4.80\times10^{-3}$} \\
\bottomrule
\end{tabular}
\end{table}
\subsection{Ablation study}
\label{sec:ablation}
To identify the sources of the predictive performance, we conduct ablation experiments on the Kolmogorov-flow dataset with $\nu=10^{-3}$ at a spatial resolution of $32\times32$. The lower resolution is chosen because full compensation requires selective-SSM scans on the high-resolution skip features: at $64\times64$ the flattened feature maps form token sequences of length $4096$, which makes SSM-based compensation at every level computationally prohibitive. The main $64\times64$ results of Section~\ref{sec:benchmark} therefore use the partial-compensation strategy, with selective-SSM modules at the bottleneck and Koopman-type linear compensation at the high-resolution skip levels. The compared variants are:
\begin{itemize}
    \item \textbf{Full compensation} (Mamba+Mamba+Mamba): selective-SSM modules at the bottleneck and at every skip level; the complete compensation scheme, affordable only at $32\times32$.
    \item \textbf{Partial compensation} (Koopman+Mamba+Mamba, variant v2): selective-SSM compensation at the bottleneck and the lowest-resolution skip levels, with Koopman-type linear compensation at the high-resolution skip levels; the configuration used in the main experiments.
    \item \textbf{All-linear compensation} (Koopman+Koopman+Koopman, variant v3): every selective-SSM module is replaced by the Koopman-type linear form, so that all compensation is state-independent; this variant isolates the role of state dependence.
    \item \textbf{No warmup}: the full model with the overlapping-warmup strategy removed, testing the role of continuous temporal context at window boundaries.
    \item \textbf{UNO T-channel}: the standard UNO with history concatenated along the channel dimension, testing whether the gain arises solely from the UNO backbone.
    \item \textbf{UNO T-batch}: history folded into the batch dimension with a shared UNO per time step, testing whether temporal reshaping alone explains the improvement.
\end{itemize}

\begin{figure}[H]
\centering
\includegraphics[width=0.86\textwidth]{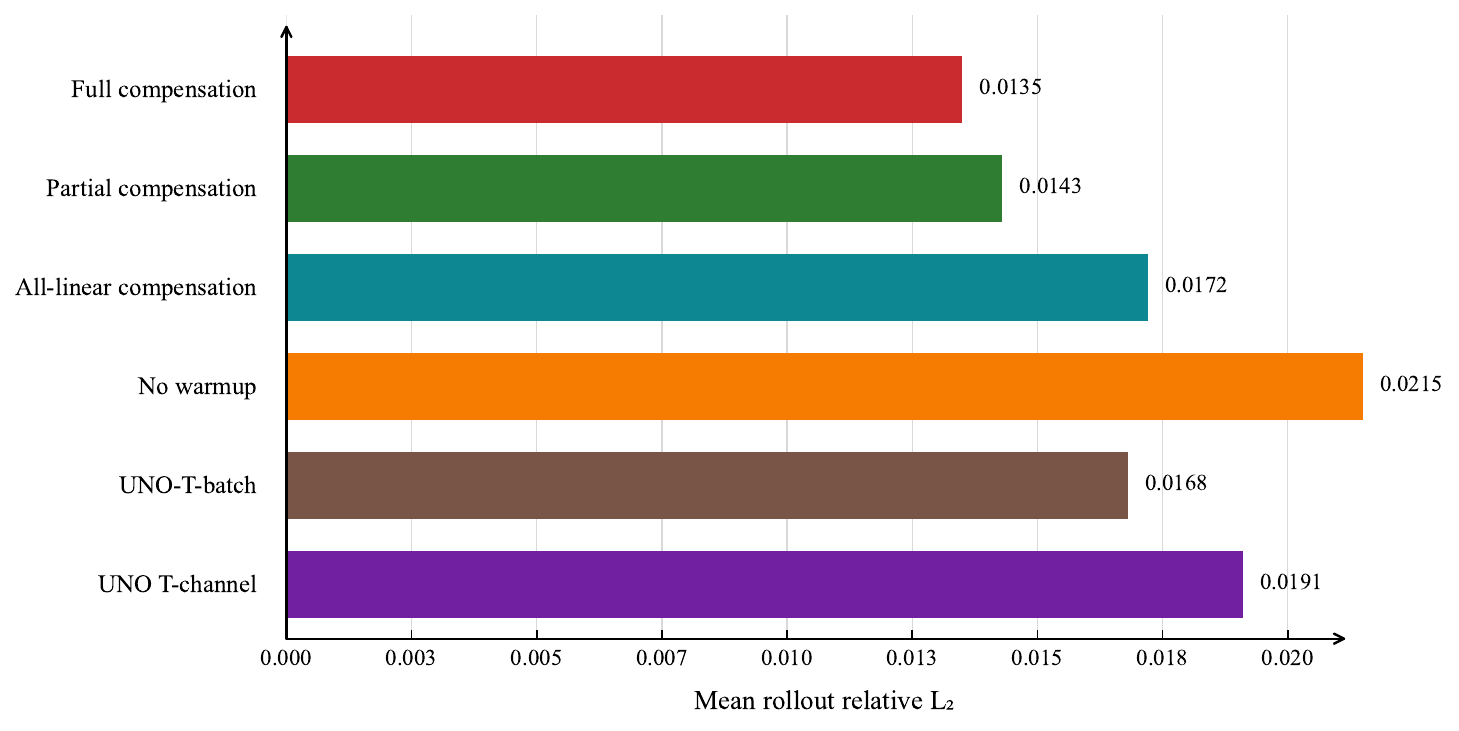}
\caption{Ablation results on Kolmogorov flow with $\nu=10^{-3}$ at $32\times32$ resolution. \NEW{Lower mean rollout relative $L_2$ error indicates better performance.}}
\label{fig:ablation}
\end{figure}

The results are shown in Fig.~\ref{fig:ablation}, and the pairwise comparisons support the following conclusions. Full vs.\ partial compensation: \NEW{additionally placing selective-SSM modules at the high-resolution skip levels reduces the error from $1.43\times10^{-2}$ to $1.35\times10^{-2}$, a $5.6\%$ reduction relative to the partial variant.} The gain is real but modest compared with its computational cost, which is why the partial strategy is used at $64\times64$. Full vs.\ no warmup: \NEW{removing warmup increases the error from $1.35\times10^{-2}$ to $2.15\times10^{-2}$; thus, the complete model reduces the error by $37.2\%$ relative to the no-warmup variant.} Without warmup, the SSM hidden state is initialized directly from self-generated predictions at each window boundary, and early errors propagate into subsequent steps. Full vs.\ UNO T-batch: \NEW{CoKo-UNO reduces the error from $1.68\times10^{-2}$ to $1.35\times10^{-2}$, a $19.6\%$ reduction,} so the improvement cannot be attributed to a mere reorganization of temporal information. UNO T-batch vs.\ UNO T-channel: folding time into the batch dimension already improves over the standard backbone, confirming that temporal structure matters. Finally, all compensated variants outperform the plain UNO backbone, confirming that multiscale spatial representation alone is insufficient. Partial vs.\ all-linear compensation: replacing the state-dependent compensation modules by purely linear propagation \NEW{increases the error from $1.43\times10^{-2}$ to $1.72\times10^{-2}$, a $20.3\%$ degradation relative to partial compensation.} This result demonstrates that the compensation must be state-dependent---a state-independent linear correction cannot represent the residual $r(x_t)$ of Eq.~\eqref{eq:koopman_residual}, which is the central premise of the proposed method. Taken together, the ablations show that the predictive capability of CoKo-UNO arises from the combination of the multiscale backbone, the selective state-space compensation of Koopman truncation residuals, and the continuous context supplied by warmup, consistent with the compensated finite-dimensional Koopman interpretation of Section~\ref{sec:method}.
\subsection{Accuracy--time trade-off}
\label{sec:tradeoff}
\begin{figure}[H]
\centering
\includegraphics[width=\textwidth]{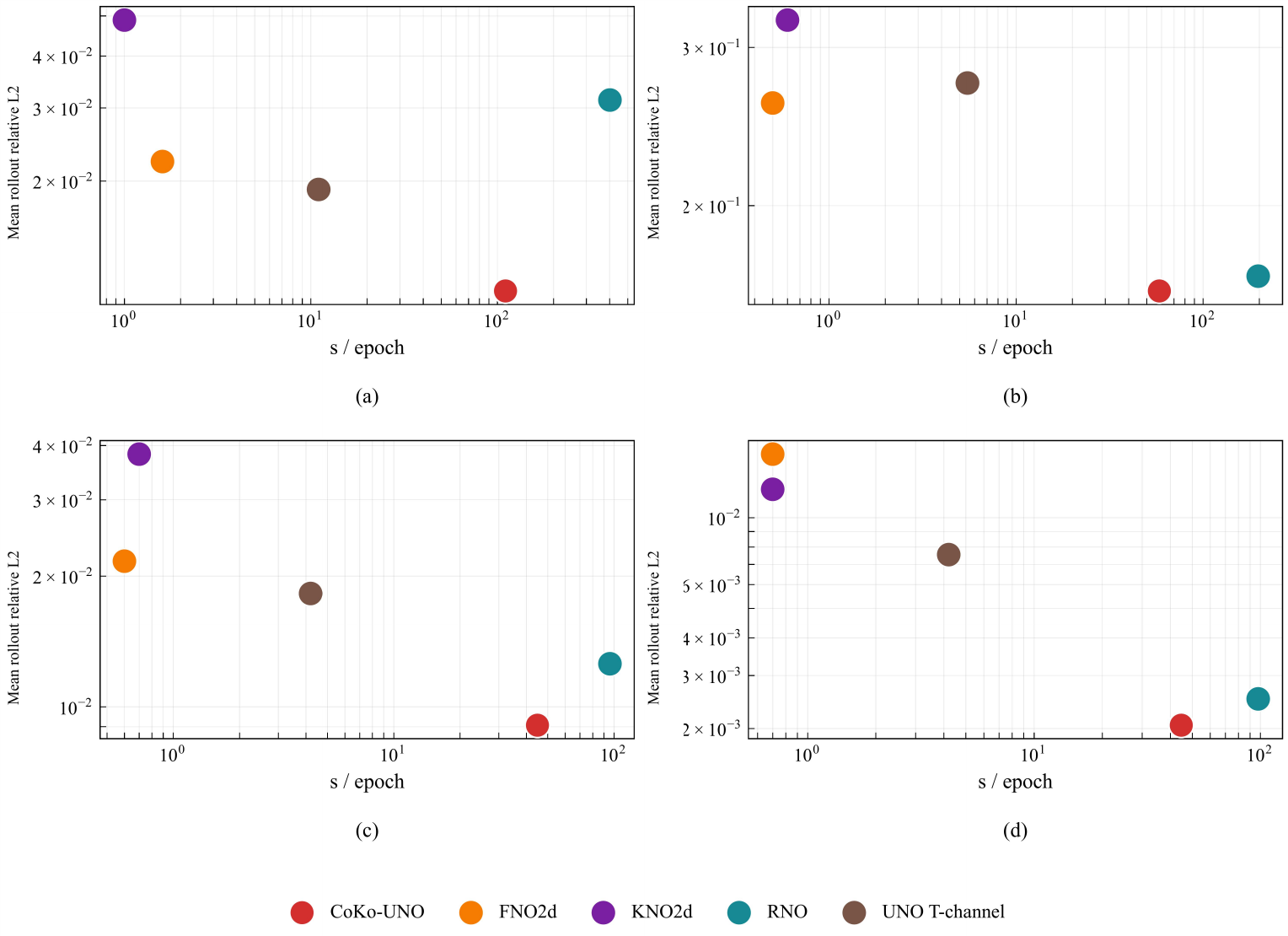}
\caption{\NEW{Accuracy--time Pareto comparison across the benchmark problems. (a) Kolmogorov flow with $\nu=10^{-3}$ at $32\times32$ resolution. (b) Kolmogorov flow with $\nu=10^{-4}$ at $32\times32$ resolution. (c) Cylinder-flow velocity magnitude $\lvert\mathbf{U}\rvert$. (d) An additional cylinder-flow field. Each point reports the mean rollout relative $L_2$ error and the average time per training epoch; lower-left placement is preferable.}}
\label{fig:accuracy_time}
\end{figure}

Beyond predictive accuracy, we evaluate computational efficiency and the resulting accuracy--time trade-off. Fig.~\ref{fig:accuracy_time} relates the \NEW{mean rollout relative $L_2$ error} to the average training time per epoch (wall-clock over epochs $1$--$199$); models closer to the lower-left corner achieve both lower errors and shorter training times. \NEW{This metric is defined in Eq.~}\eqref{eq:mean_rel_l2}\NEW{ and is used consistently for the accuracy comparisons in Fig.~}\ref{fig:accuracy_time}\NEW{ and Table~}\ref{tab:main_results}\NEW{.}

\begin{table}[H]
\centering
\caption{Average training time per epoch and total training time over $200$ epochs for the ablation variants and baseline methods. \NEW{Entries are reported as mean $\pm$ standard deviation over three independent runs.}}
\label{tab:efficiency}
\begin{tabular}{lcc}
\toprule
Method & Average time per epoch & Total time for $200$ epochs \\
\midrule
FNO2d          & \NEW{$2.229\pm0.005\,\mathrm{s}$} & \NEW{$0.124\pm0.0003\,\mathrm{h}$} \\
KNO2d          & \NEW{$2.395\pm0.012\,\mathrm{s}$} & \NEW{$0.133\pm0.0007\,\mathrm{h}$} \\
RNO            & \NEW{$392.99\pm0.67\,\mathrm{s}$} & \NEW{$21.83\pm0.04\,\mathrm{h}$} \\
UNO T-channel  & \NEW{$16.80\pm0.07\,\mathrm{s}$} & \NEW{$0.933\pm0.004\,\mathrm{h}$} \\
UNO T-batch    & \NEW{$161.60\pm0.34\,\mathrm{s}$} & \NEW{$8.98\pm0.02\,\mathrm{h}$} \\
CoKo-UNO       & \NEW{$162.71\pm0.05\,\mathrm{s}$} & \NEW{$9.039\pm0.003\,\mathrm{h}$} \\
\bottomrule
\end{tabular}
\end{table}

As reported in Table~\ref{tab:efficiency} and visualized in Fig.~\ref{fig:accuracy_time}, FNO2d and KNO2d have the lowest per-epoch costs, but their rollout errors are substantially higher than those of models with explicit temporal modeling on most benchmark cases: their low cost mainly reflects simple temporal processing, which cannot control error accumulation in extended autoregressive rollout. UNO T-channel is likewise inexpensive but noticeably less accurate than CoKo-UNO, indicating that the multiscale backbone alone is insufficient for stable long-term prediction.

RNO is the strongest temporal baseline but also the most expensive one: on Kolmogorov flow with $\nu=10^{-3}$, its average epoch time is \NEW{$392.99\pm0.67\,\mathrm{s}$}, compared with \NEW{$162.71\pm0.05\,\mathrm{s}$} for CoKo-UNO, a reduction of about \NEW{$58.60\%$}; at the same time, CoKo-UNO improves the \NEW{mean rollout relative} $L_2$ \NEW{error from $1.27\times10^{-2}$ to $1.09\times10^{-2}$.} Across the remaining tasks, the average epoch time of CoKo-UNO is typically only about \NEW{$41.40\%$} of that of RNO, while achieving lower rollout errors. Compared with a fully recurrent neural operator, CoKo-UNO therefore offers a far more favorable balance between rollout accuracy and computational cost.

UNO T-batch has \NEW{a training time nearly identical to that of} CoKo-UNO and achieves competitive accuracy on the short-horizon cylinder-flow cases, indicating that for relatively regular dynamics over short horizons, simple temporal reshaping can be an effective solution. On the two Kolmogorov-flow cases, however, CoKo-UNO achieves more stable accuracy and better control of error growth during extended rollout, which justifies its \NEW{small} additional cost.

Overall, CoKo-UNO is not the least expensive model, but its additional computation translates into substantial improvements in rollout accuracy: it provides stronger temporal modeling than FNO2d, KNO2d, and UNO T-channel, and it is simultaneously more accurate and about \NEW{$58.60\%$} cheaper to train than the fully recurrent RNO. The accuracy--time Pareto results therefore show that selective state-space compensation improves autoregressive prediction without incurring the computational burden of a fully recurrent architecture.

\subsection{Physical diagnostics}
\label{sec:diagnostics}
Beyond field-wise errors, we assess whether the rollouts preserve physically meaningful structure: we first inspect the predicted fields directly, and then examine the kinetic-energy level and energy spectrum for Kolmogorov flow, water-mass conservation for SWE, and wake-region fidelity for cylinder flow.

\begin{figure}[H]
\centering
\includegraphics[width=0.68\textwidth]{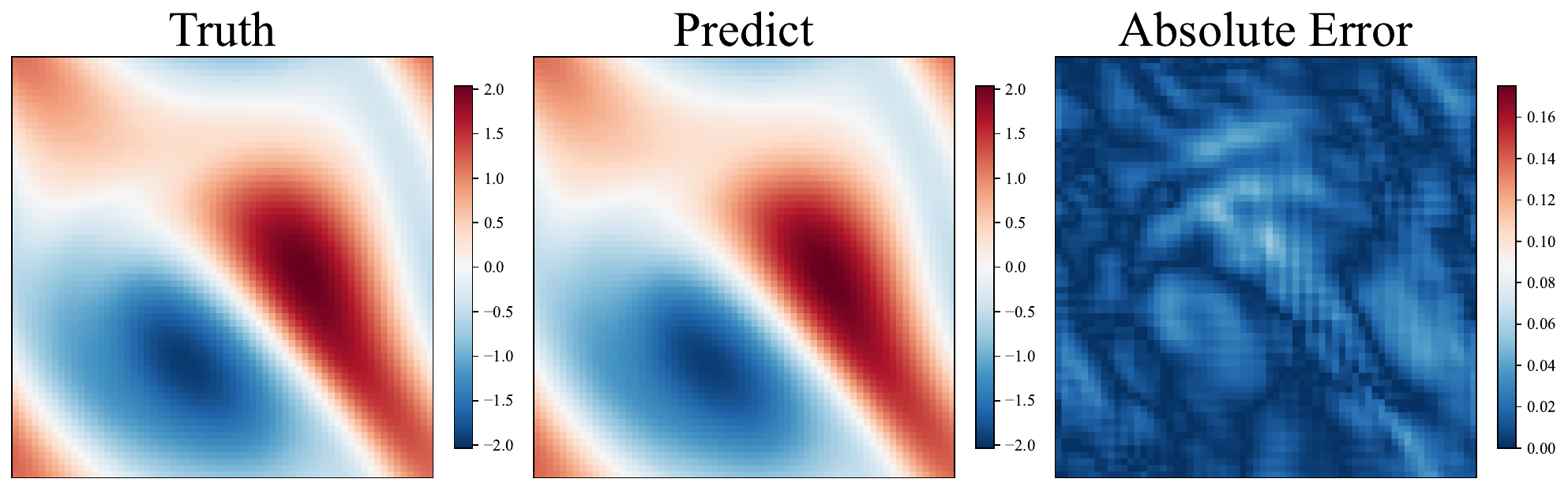}\\[-0.35em]
{\footnotesize\textnormal{(a)}}\\[0.35em]
\includegraphics[width=0.68\textwidth]{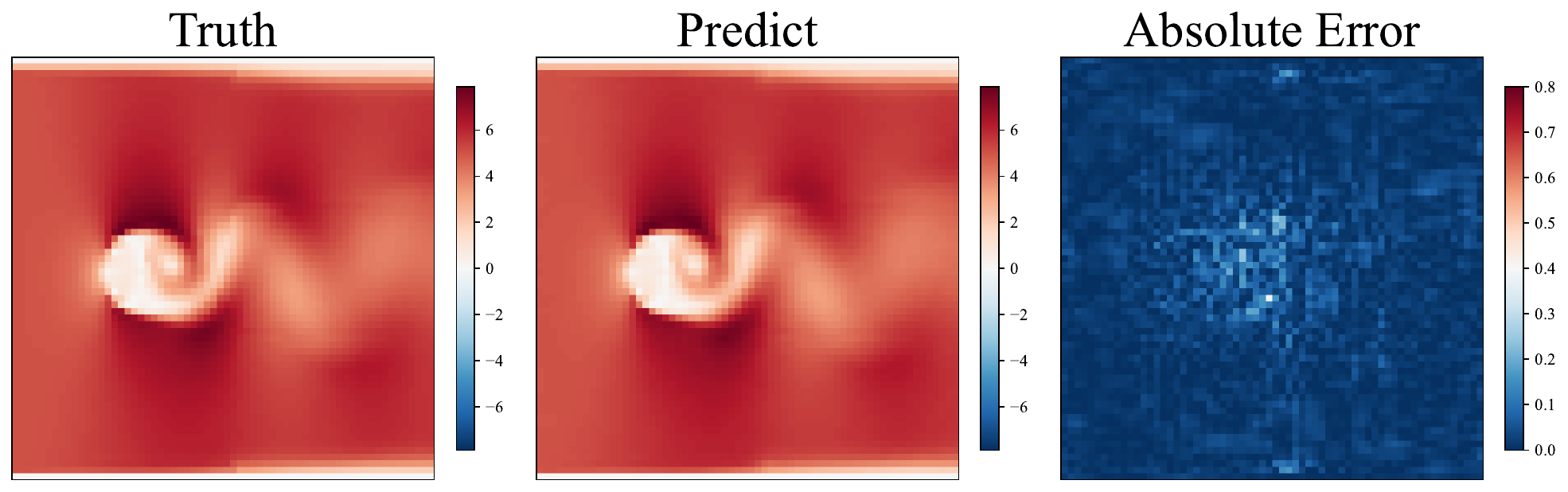}\\[-0.35em]
{\footnotesize\textnormal{(b)}}\\[0.35em]
\includegraphics[width=0.68\textwidth]{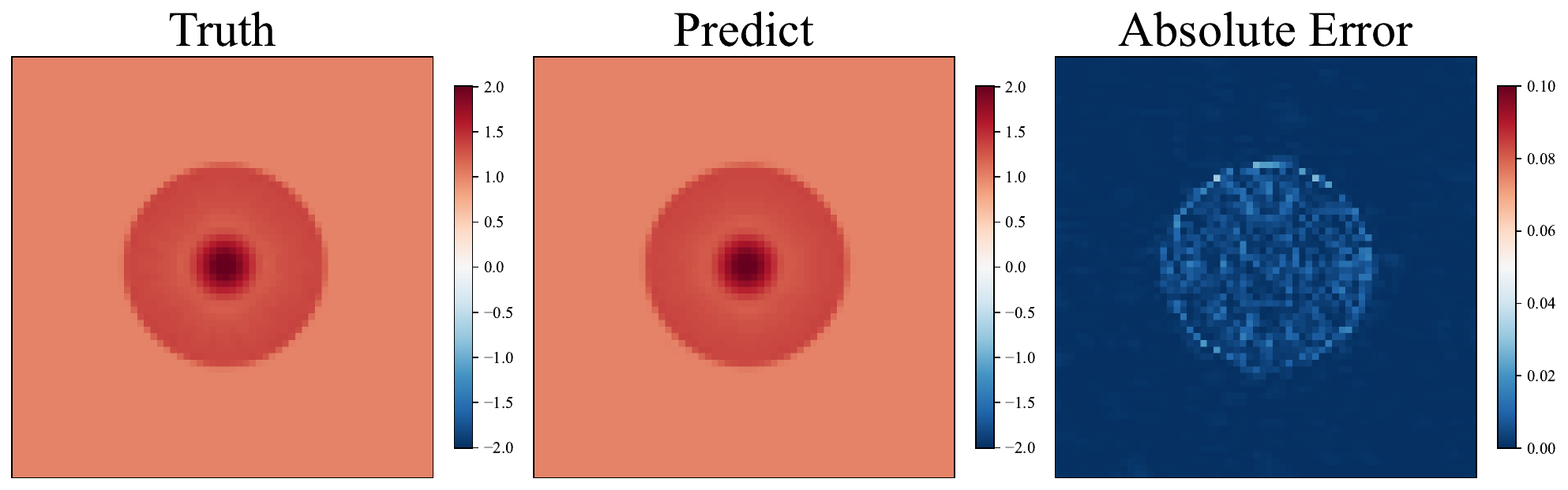}\\[-0.35em]
{\footnotesize\textnormal{(c)}}
\caption{\NEW{Representative CoKo-UNO predictions at a late rollout step. Each subfigure shows the ground-truth field, the prediction, and the absolute error: (a) Kolmogorov-flow vorticity at $\nu=10^{-3}$, (b) cylinder-flow velocity magnitude $\lvert\mathbf{U}\rvert$, and (c) water level for the shallow-water equations.}}
\label{fig:field_snapshots}
\end{figure}

\NEW{Figure}~\ref{fig:field_snapshots} compares representative predicted fields with the reference solution for all three flow problems at a late rollout step. The large-scale vortex structures, the cylinder wake pattern, and the shallow-water wave fronts remain coherent throughout the rollout, and the pointwise error is localized at thin vortex filaments, shear layers, and wave fronts rather than being spread over the domain. Corresponding visualizations for the baseline models, which exhibit visibly stronger smearing and phase drift, are collected in the appendix.
\subsubsection{Kolmogorov flow: kinetic energy and energy spectrum}
For the forced Kolmogorov-flow cases, we evaluate the domain-mean kinetic energy and the time-averaged kinetic-energy spectrum \cite{pope2000turbulent}. For a velocity field $\boldsymbol{u}=(u,v)$, the former is
\begin{equation}
    K(t)=\frac{1}{2|\Omega|}\int_{\Omega}\left(u^2(\boldsymbol{x},t)+v^2(\boldsymbol{x},t)\right)\,\mathrm{d}\boldsymbol{x},
\end{equation}
and the latter is the shell-averaged Fourier energy $E(k)$. These measures complement the pointwise vorticity error: $K(t)$ tests the global energetic level, whereas $E(k)$ tests whether energy is placed at the correct spatial scales. Because the NS cases are forced, the kinetic-energy curves are interpreted as an evolution towards a statistical level rather than as a monotone decay law.

\begin{figure}[H]
\centering
\includegraphics[width=0.98\textwidth]{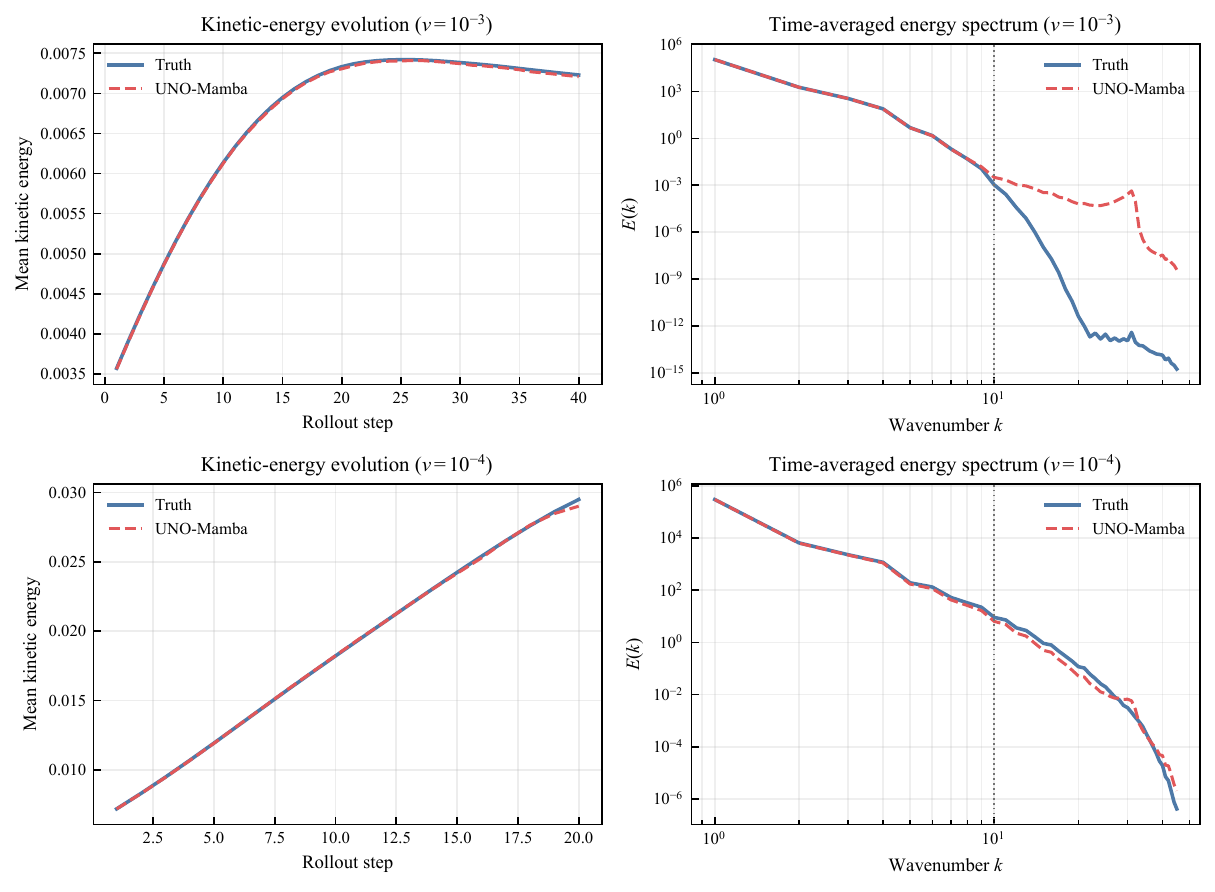}
\caption{Physical diagnostics of the selected proposed-model runs on forced Kolmogorov-flow data. Left: mean kinetic-energy evolution. Right: time-averaged energy spectrum; the dotted line marks $k=10$.}
\label{fig:app_ns_physics}
\end{figure}

\NEW{The diagnostics shown in Fig.~}\ref{fig:app_ns_physics}\NEW{ are computed directly from the saved rollout data. For $\nu=10^{-3}$, the mean relative discrepancy of $K(t)$ is $2.10\times10^{-3}$ over the rollout (maximum $3.60\times10^{-3}$ and final-step value $3.00\times10^{-3}$). For the more demanding $\nu=10^{-4}$ case, the corresponding mean discrepancy remains $2.40\times10^{-3}$; it reaches $1.63\times10^{-2}$ only at the final step. Thus, the low vorticity error is accompanied by close preservation of the global energetic level, rather than being caused by a uniformly damped forecast. The spectra also agree closely in the resolved range $1\leq k\leq10$: the mean relative spectral discrepancies are $24.30\%$ and $13.10\%$ for $\nu=10^{-3}$ and $\nu=10^{-4}$, respectively. Differences become visibly larger beyond the marked cutoff, where spectral energy is very small and relative ratios are correspondingly ill-conditioned. This localization of the mismatch to weak, high-wavenumber content is consistent with finite-resolution autoregressive prediction and does not alter the accurately reproduced large-scale energy distribution.}

\subsubsection{\NEW{Cylinder flow: wake-region velocity-magnitude error}}
\NEW{Wake fidelity is assessed separately from the full-domain velocity-magnitude error, since the downstream recirculation and vortex-shedding region is the most sensitive portion of cylinder flow} \cite{williamson1996vortex}\NEW{. Let $\Omega_{\mathrm{wake}}$ denote the prescribed downstream wake mask. We compute, at each rollout step,}
\begin{equation}
    \varepsilon_{\mathrm{wake}}(t)=
    \frac{\left\|\widehat{\lvert\mathbf{U}\rvert}(t)-\lvert\mathbf{U}\rvert(t)\right\|_{2,\Omega_{\mathrm{wake}}}}
    {\left\|\lvert\mathbf{U}\rvert(t)\right\|_{2,\Omega_{\mathrm{wake}}}}.
\end{equation}
For $N$ saved rollout samples, the plotted mean and its normal-approximation 95\% confidence interval are
\begin{equation}
    \bar{\varepsilon}_{\mathrm{wake}}(t)=\frac{1}{N}\sum_{i=1}^{N}\varepsilon_{\mathrm{wake}}^{(i)}(t),
    \qquad
    \mathrm{CI}_{95\%}(t)=\bar{\varepsilon}_{\mathrm{wake}}(t)\pm1.96\frac{s_{\mathrm{wake}}(t)}{\sqrt{N}},
\end{equation}
where $s_{\mathrm{wake}}(t)$ is the sample standard deviation of $\varepsilon_{\mathrm{wake}}^{(i)}(t)$ across the saved rollouts.

\begin{figure}[H]
\centering
\includegraphics[width=0.68\textwidth]{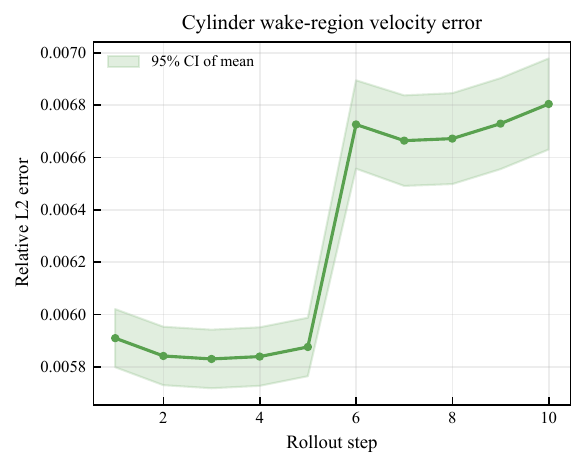}
\caption{\NEW{Cylinder wake-region relative velocity-magnitude error. Shading denotes the normal-approximation 95\% confidence interval of the mean across saved rollout samples.}}
\label{fig:app_cylinder_wake}
\end{figure}

\NEW{Figure}~\ref{fig:app_cylinder_wake} \NEW{shows a mean wake-region velocity-magnitude error of approximately $6.30\times10^{-3}$ across the saved rollout samples. It remains near $5.80$--$5.90\times10^{-3}$ during the first five steps, then rises to about $6.70$--$6.80\times10^{-3}$ for the remaining rollout steps. Hence, error accumulation remains mild precisely in the wake, where small phase or amplitude errors can otherwise quickly distort the vortex street. Together with the lower full-field cylinder velocity-magnitude error in Table~}\ref{tab:main_results}\NEW{, this diagnostic indicates that the improvement is retained in the dynamically important downstream region rather than being confined to the nearly uniform upstream flow.}

\subsubsection{SWE: relative water-mass drift}
For the shallow-water rollout, we evaluate the water mass, a conserved integral quantity for the continuous system \cite{leveque2002finite},
\begin{equation}
    M_h(t)=\int_{\Omega} h(\boldsymbol{x},t)\,\mathrm{d}\boldsymbol{x},
    \qquad
    \varepsilon_M(t)=
    \frac{\left|\widehat{M}_h(t)-M_h(t)\right|}{\left|M_h^0\right|},
\end{equation}
where $h$ is water depth and $M_h^0$ is the reference water mass used for normalization. On the uniform $64\times64$ grid, the integral is evaluated as $M_h\approx\sum_{i,j}h_{ij}\,\Delta x\Delta y$. Unlike a field-wise relative $L_2$ error, $\varepsilon_M$ detects accumulated bias in the conserved bulk quantity while remaining independent of the absolute mass scale.

For $N$ saved rollout samples, the plotted mean drift and its normal-approximation 95\% confidence interval are
\begin{equation}
    \bar{\varepsilon}_M(t)=\frac{1}{N}\sum_{i=1}^{N}\varepsilon_M^{(i)}(t),
    \qquad
    \mathrm{CI}_{95\%}(t)=\bar{\varepsilon}_M(t)\pm1.96\frac{s_M(t)}{\sqrt{N}},
\end{equation}
where $s_M(t)$ is the sample standard deviation of $\varepsilon_M^{(i)}(t)$ across the saved rollouts.

\begin{figure}[H]
\centering
\includegraphics[width=0.68\textwidth]{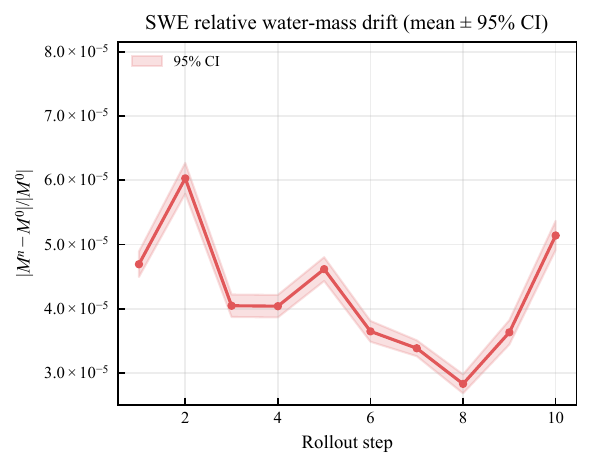}
\caption{Relative SWE water-mass drift, normalized by the reference mass $M_h^0$. The dashed line marks zero drift, and shading denotes the normal-approximation 95\% confidence interval of the mean across saved rollout samples.}
\label{fig:app_swe_mass}
\end{figure}

As shown in Fig.~\ref{fig:app_swe_mass}, the mean relative water-mass drift is approximately $4.20\times10^{-5}$ over the rollout. It starts at about $4.70\times10^{-5}$, reaches a maximum of $6.00\times10^{-5}$ at the second rollout step, and remains below $5.20\times10^{-5}$ thereafter; the final-step value is about $5.10\times10^{-5}$. Thus, the drift remains on the order of $10^{-5}$ and shows no monotonic accumulation over the ten-step horizon. Although no hard conservation constraint is imposed during training, this bounded relative drift indicates that recursive prediction does not introduce a substantial systematic water-mass bias.

\NEW{\paragraph{Resolution-transfer assessment.} To assess spatial-resolution generalization, we train the models on Kolmogorov flow with $\nu=10^{-3}$ at $32\times32$ resolution and directly evaluate them, without fine-tuning, at $32\times32$, $64\times64$, and $128\times128$. The results in Table~}\ref{tab:resolution_transfer}\NEW{ show that the standard UNO T-channel baseline remains relatively stable at the training resolution but deteriorates at the two higher resolutions, whereas CoKo-UNO also shows a substantial increase in error as the evaluation resolution departs from the training resolution: its global relative $L_2$ error rises from $1.43\times10^{-2}$ at $32\times32$ to $2.87\times10^{-1}$ and $4.21\times10^{-1}$ at $64\times64$ and $128\times128$, respectively. CoKo-UNO performs slightly better than UNO T-channel at $32\times32$ and $128\times128$, while its error is marginally higher at $64\times64$. This behavior is consistent with the resolution-dependent latent compensation and skip pathways: the present implementation couples the learned compensation to the training grid, which weakens direct resolution transfer. Accordingly, these Kolmogorov-flow results should not be interpreted as evidence of zero-shot resolution generalization for CoKo-UNO; developing resolution-decoupled compensation and evaluating it on additional operator-learning tasks remain important future work.}

\section{Conclusion}\label{sec:conclusion}
We introduced CoKo-UNO, a Compensated Koopman U-shaped Neural Operator for stable autoregressive prediction of unsteady flows. The starting point of this work is the observation that finite-dimensional Koopman models leave a state-dependent truncation residual that is repeatedly injected and propagated during autoregressive rollout, \NEW{as described by Eq.~}\eqref{eq:error_recursion}\NEW{.} CoKo-UNO compensates this residual explicitly: a selective state-space model, whose input-dependent update is shown to be the natural parameterization of the compensation system, corrects the dominant linear propagation at the bottleneck of a multiscale spectral U-shaped backbone, and the compensation is transmitted upward through resolution-adaptive compensatory skip connections, while an overlapping-warmup strategy provides continuous context at rollout-window boundaries.

\NEW{On four benchmark problems---Kolmogorov flow at two viscosities, cylinder flow, and the shallow-water equations---CoKo-UNO achieves the lowest mean rollout error among all compared neural operators: it reduces the mean error by $85.71\%$ relative to KNO2d on Kolmogorov flow at $\nu=10^{-3}$, and by up to $76.76\%$ relative to the strongest baseline while requiring about $41.40\%$ of RNO's training cost.} Step-wise error curves and the fitted growth rates show that the advantage stems from slower error accumulation rather than better one-step fitting, and the physical diagnostics (kinetic energy, energy spectra, water-mass conservation, wake-region error) together with the flow-field visualizations confirm that the rollouts preserve the physically relevant structure of the flow.

Several limitations remain. First, the gains are limited where the strongest baselines are already very accurate, as on cylinder flow and SWE, and where the error is dominated by one-step representation bias rather than by its accumulation, as on Kolmogorov flow at $\nu=10^{-4}$; combining compensation with more accurate spatial representations for high-Reynolds-number flows is a natural next step. Second, full compensation at every skip level requires selective-SSM scans on long token sequences and is computationally prohibitive at resolutions beyond $64\times64$; more selective or data-efficient compensation schemes are needed for high-resolution applications. Finally, future work will investigate higher-dimensional systems, adaptive stability constraints for the dominant propagation operator, and data-efficient compensation mechanisms.

\section*{Statement}
During the preparation of this work, the authors used ChatGPT in order to improve readability and language. After using this tool/service, the authors reviewed and edited the content as needed and take full responsibility for the content of the publication.

This research was supported by the Jiangxi Provincial Natural Science Foundation (No. S20254688).

\section*{Author Contributions}
CRediT: Lv Tangying: Conceptualization, Data curation, Formal analysis, Investigation, Methodology, Software, Validation, Visualization, Writing--original draft; Dai Yuanjun: Conceptualization, Funding acquisition, Project administration, Supervision, Writing--review \& editing; Sun Zhenxu: Supervision, Writing--review \& editing.

\appendix
\section{Resolution-transfer results}
\label{app:resolution_transfer}
\NEW{Table~}\ref{tab:resolution_transfer}\NEW{ reports the zero-shot resolution-transfer results discussed at the end of Section~}\ref{sec:experiments}\NEW{.}

\begin{table}[H]
\centering
\caption{\NEW{Zero-shot resolution-transfer results on Kolmogorov flow with $\nu=10^{-3}$. Models are trained at $32\times32$ resolution and evaluated directly at the listed resolutions. Entries are global relative $L_2$ errors; lower is better.}}
\label{tab:resolution_transfer}
\begin{tabular}{lccc}
\toprule
Method & $32\times32$ $L_2\downarrow$ & $64\times64$ $L_2\downarrow$ & $128\times128$ $L_2\downarrow$ \\
\midrule
UNO T-channel & $1.91\times10^{-2}$ & $2.84\times10^{-1}$ & $4.55\times10^{-1}$ \\
CoKo-UNO & $1.43\times10^{-2}$ & $2.87\times10^{-1}$ & $4.21\times10^{-1}$ \\
\bottomrule
\end{tabular}
\end{table}

\section{Additional flow-field visualizations}
\label{app:visualizations}
\begin{figure}[H]
\centering
\includegraphics[width=0.68\textwidth]{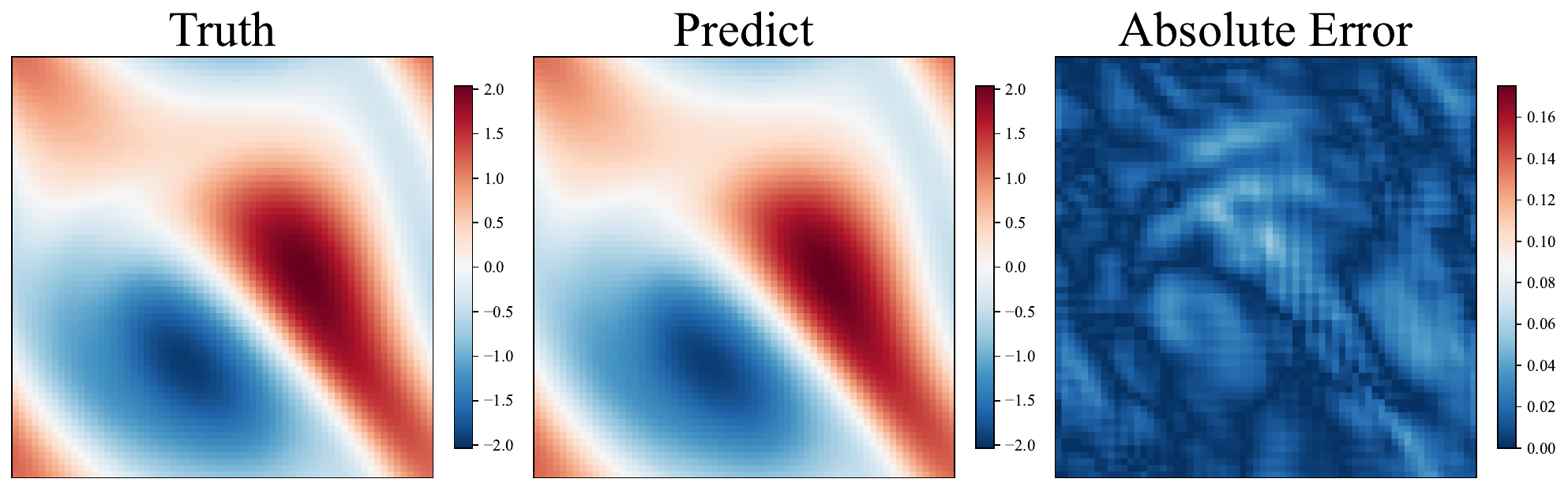}\\[-0.35em]
{\footnotesize\textnormal{(a)}}\\[0.3em]
\includegraphics[width=0.68\textwidth]{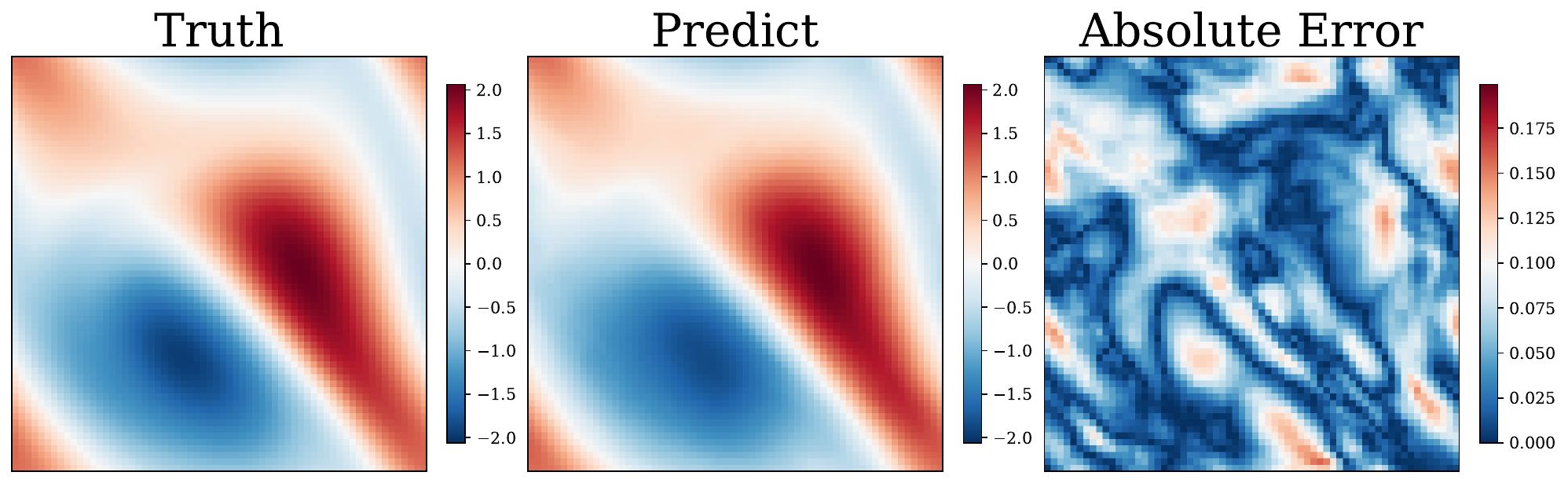}\\[-0.35em]
{\footnotesize\textnormal{(b)}}\\[0.3em]
\includegraphics[width=0.68\textwidth]{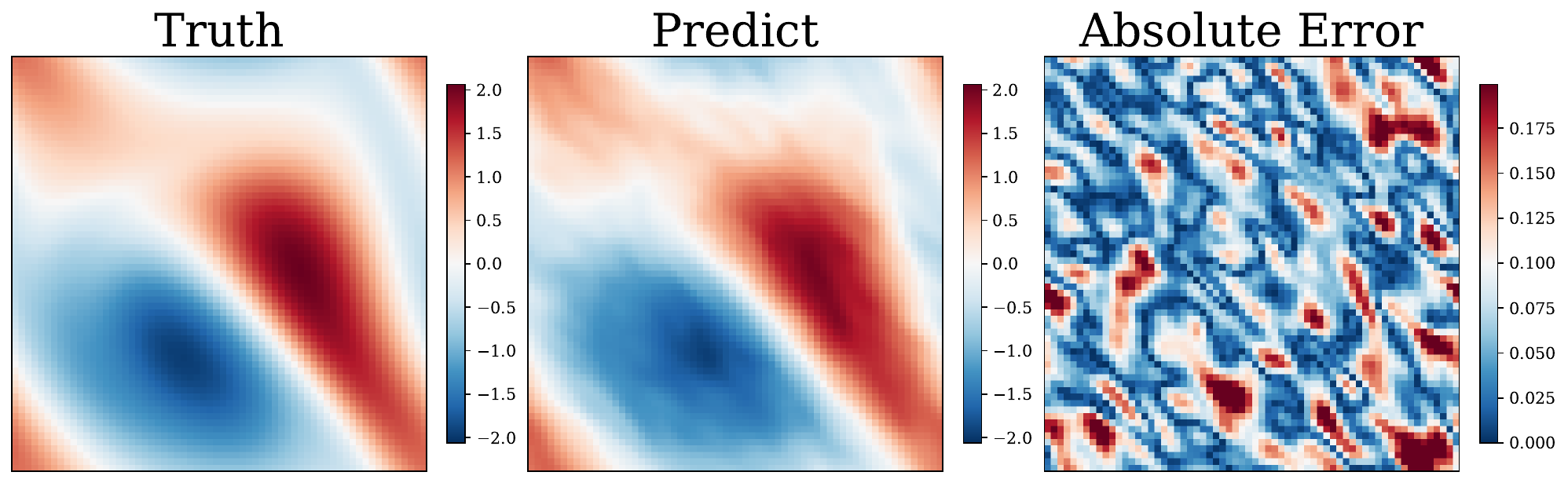}\\[-0.35em]
{\footnotesize\textnormal{(c)}}\\[0.3em]
\includegraphics[width=0.68\textwidth]{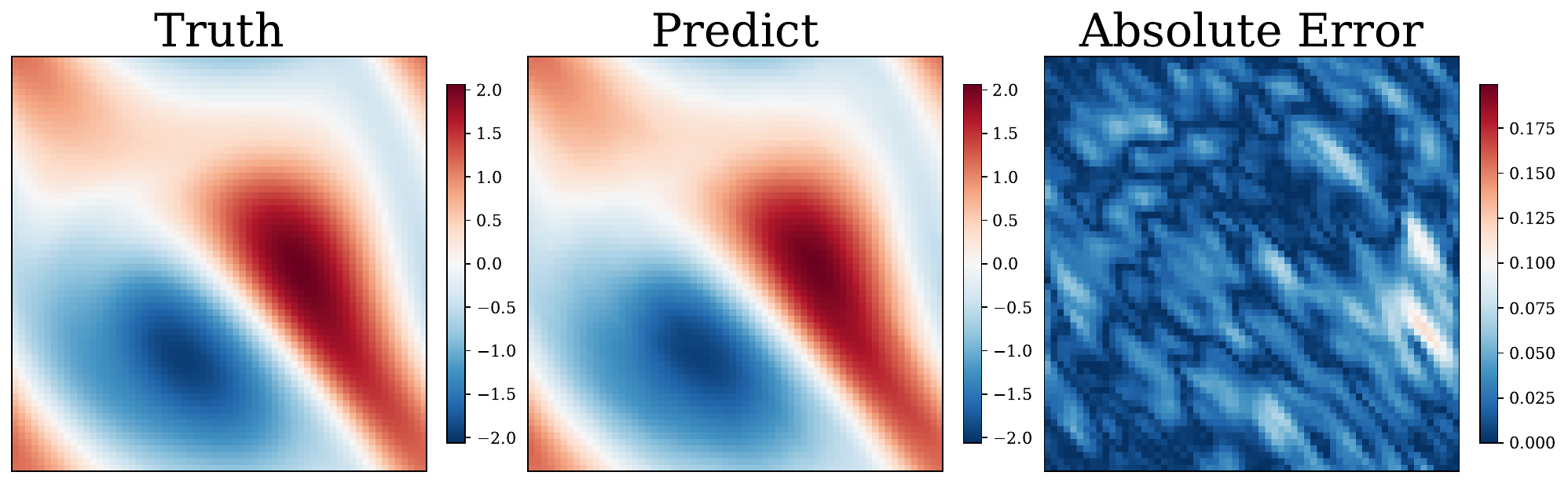}\\[-0.35em]
{\footnotesize\textnormal{(d)}}\\[0.3em]
\includegraphics[width=0.68\textwidth]{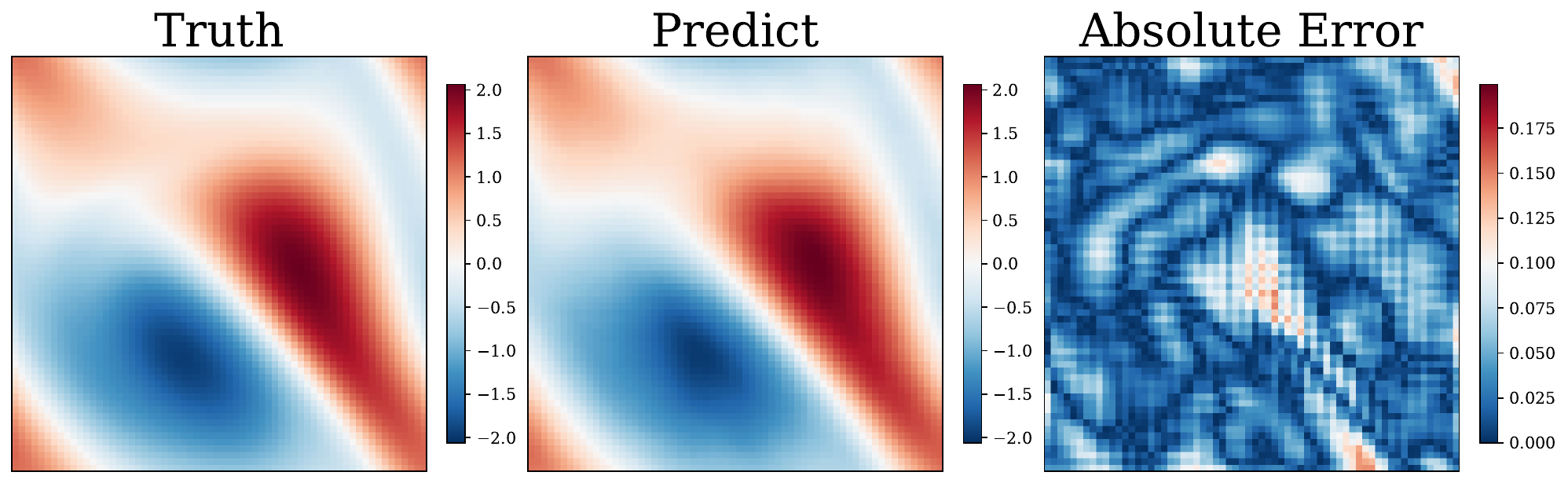}\\[-0.35em]
{\footnotesize\textnormal{(e)}}
\caption{Kolmogorov-flow vorticity predictions at $\nu=10^{-3}$. Each subfigure compares the ground-truth field (left), the predicted field (center), and the pointwise absolute error (right): (a) CoKo-UNO, (b) FNO2d, (c) KNO2d, (d) RNO, and (e) \NEW{UNO T-channel}.}
\label{fig:app_kolmogorov_fields}
\end{figure}

\begin{figure}[H]
\centering
\includegraphics[width=0.68\textwidth]{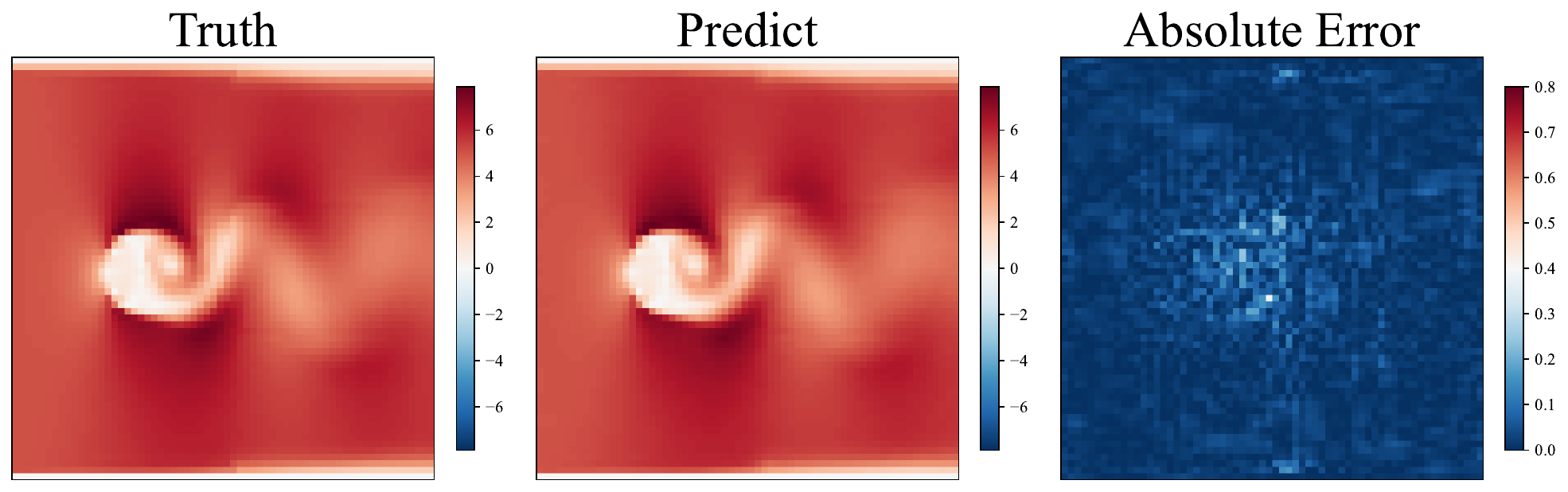}\\[-0.35em]
{\footnotesize\textnormal{(a)}}\\[0.3em]
\includegraphics[width=0.68\textwidth]{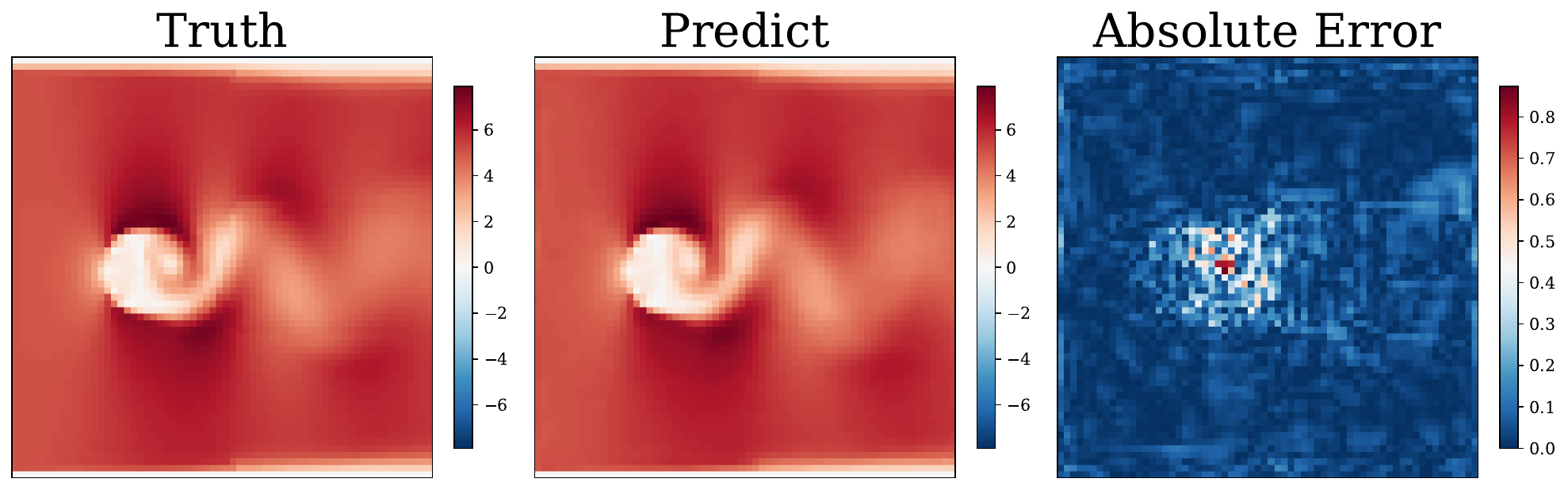}\\[-0.35em]
{\footnotesize\textnormal{(b)}}\\[0.3em]
\includegraphics[width=0.68\textwidth]{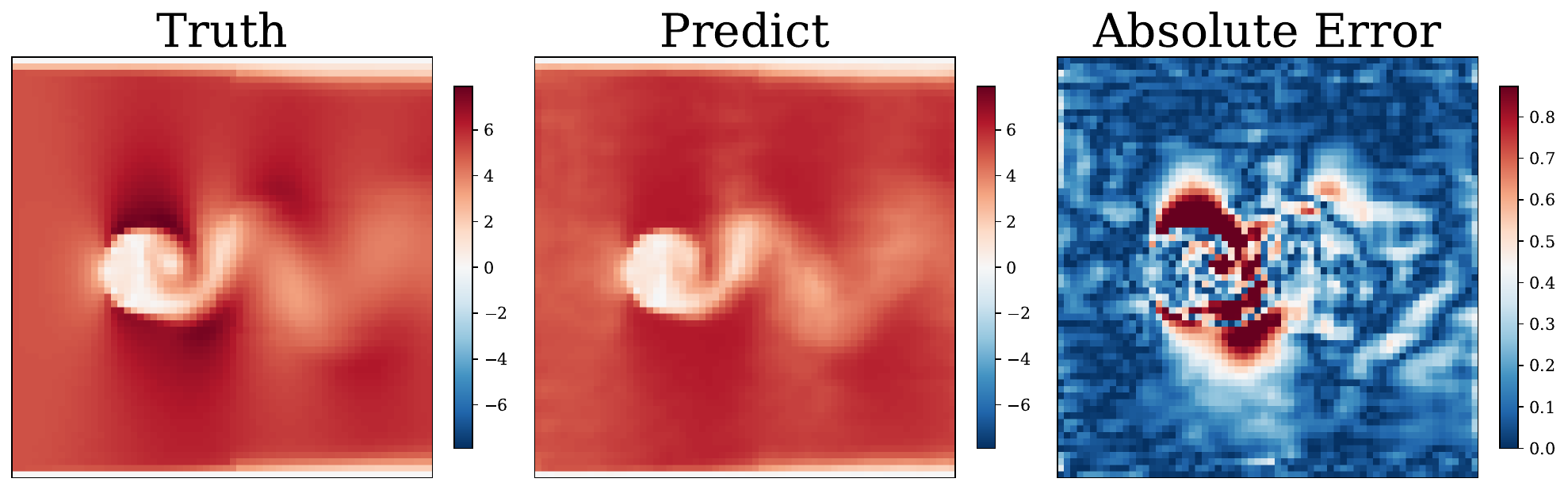}\\[-0.35em]
{\footnotesize\textnormal{(c)}}\\[0.3em]
\includegraphics[width=0.68\textwidth]{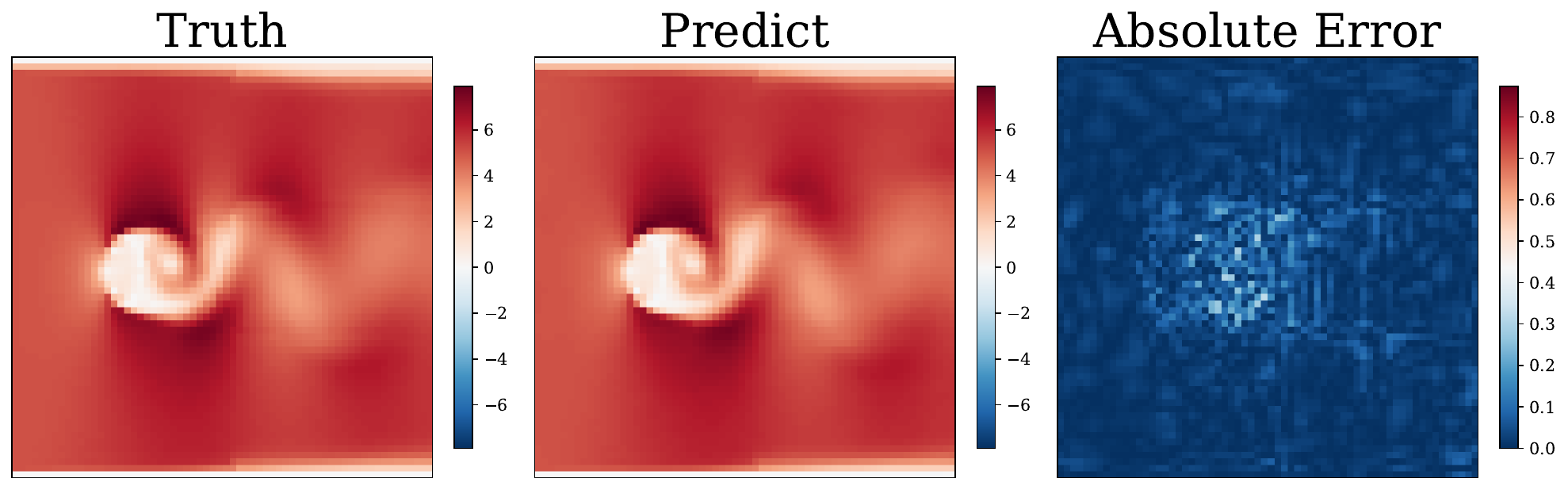}\\[-0.35em]
{\footnotesize\textnormal{(d)}}\\[0.3em]
\includegraphics[width=0.68\textwidth]{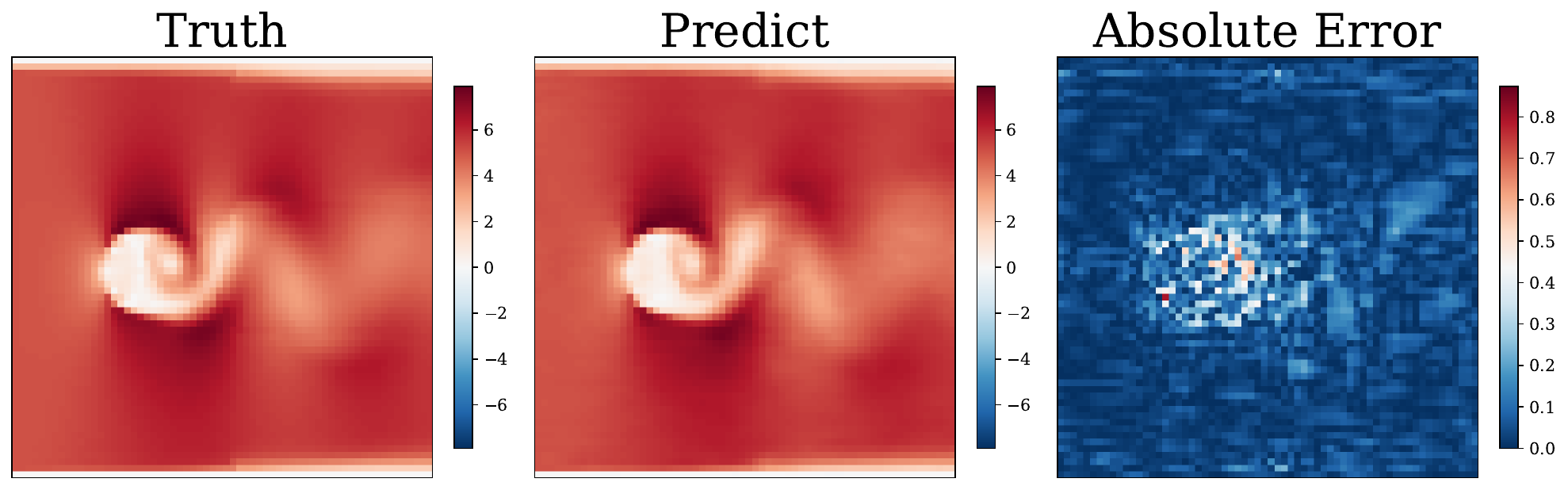}\\[-0.35em]
{\footnotesize\textnormal{(e)}}
\caption{Velocity-magnitude predictions for flow past a cylinder. Each subfigure compares the ground-truth field (left), the predicted field (center), and the pointwise absolute error (right): (a) CoKo-UNO, (b) FNO2d, (c) KNO2d, (d) RNO, and (e) \NEW{UNO T-channel}.}
\label{fig:app_cylinder_fields}
\end{figure}

\begin{figure}[H]
\centering
\includegraphics[width=0.68\textwidth]{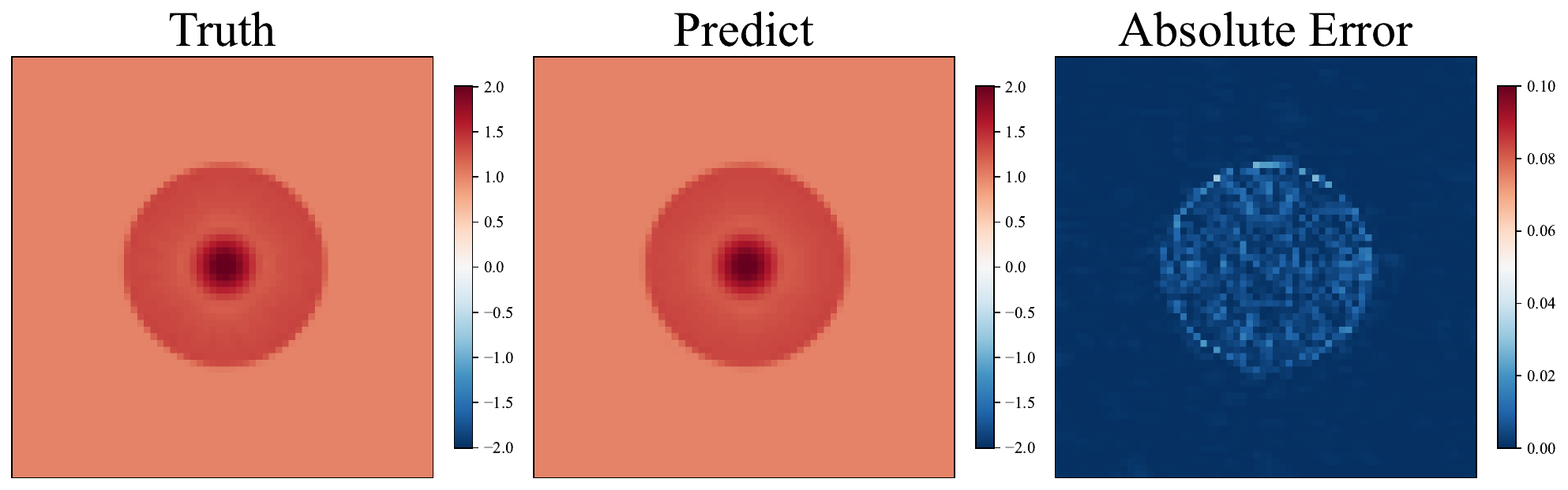}\\[-0.35em]
{\footnotesize\textnormal{(a)}}\\[0.3em]
\includegraphics[width=0.68\textwidth]{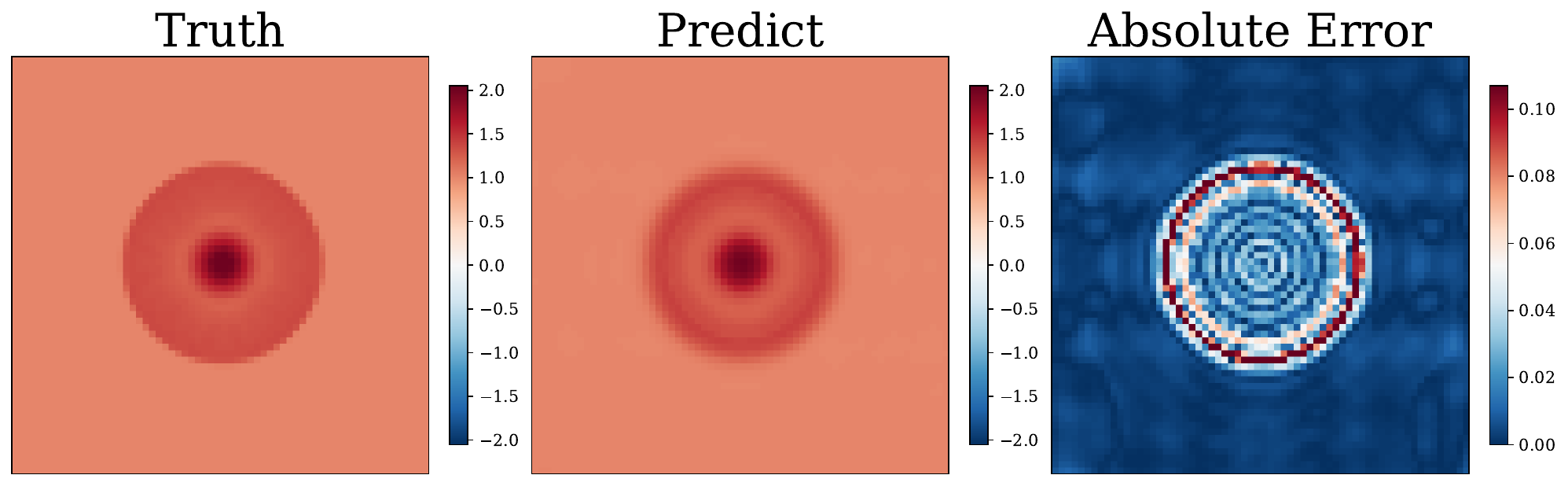}\\[-0.35em]
{\footnotesize\textnormal{(b)}}\\[0.3em]
\includegraphics[width=0.68\textwidth]{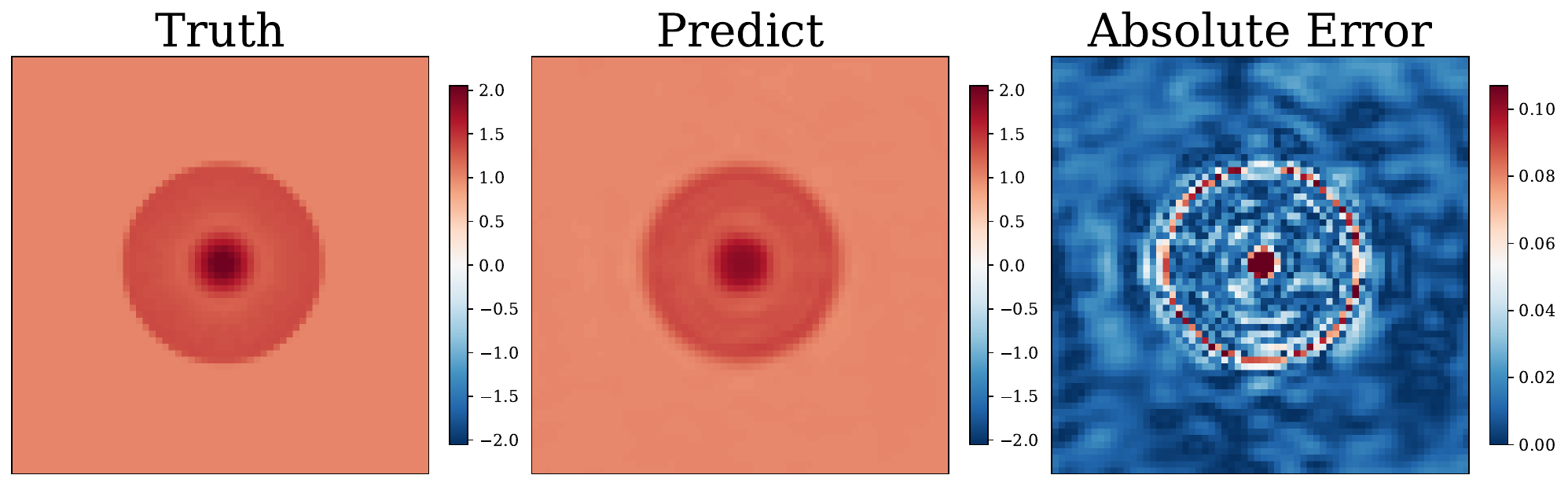}\\[-0.35em]
{\footnotesize\textnormal{(c)}}\\[0.3em]
\includegraphics[width=0.68\textwidth]{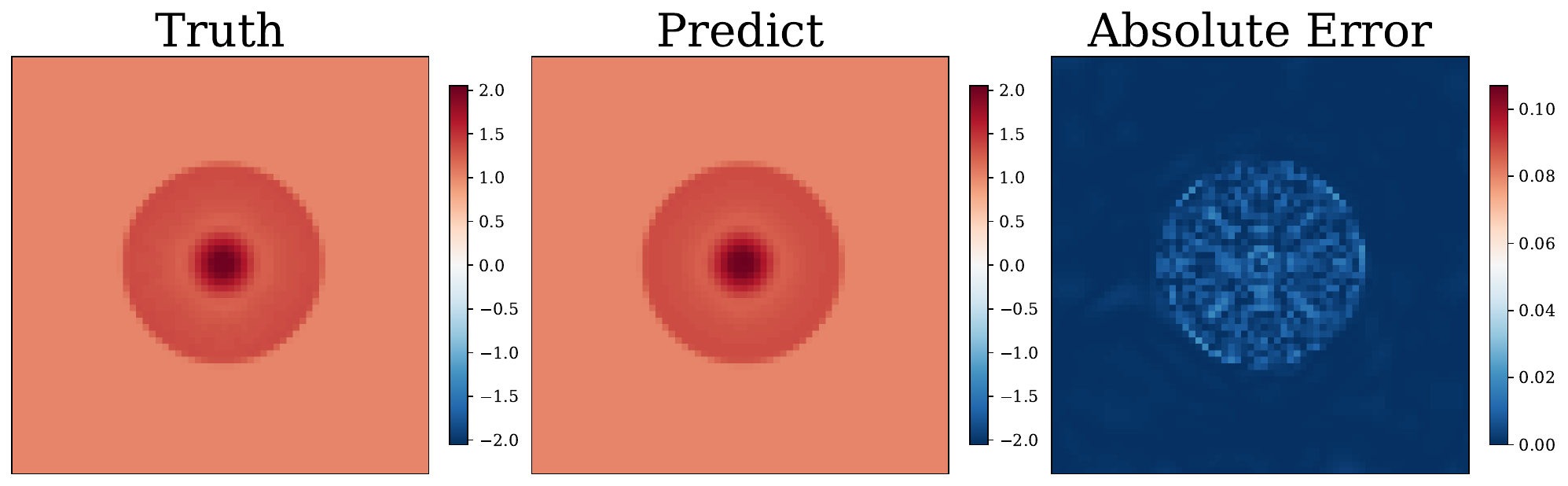}\\[-0.35em]
{\footnotesize\textnormal{(d)}}\\[0.3em]
\includegraphics[width=0.68\textwidth]{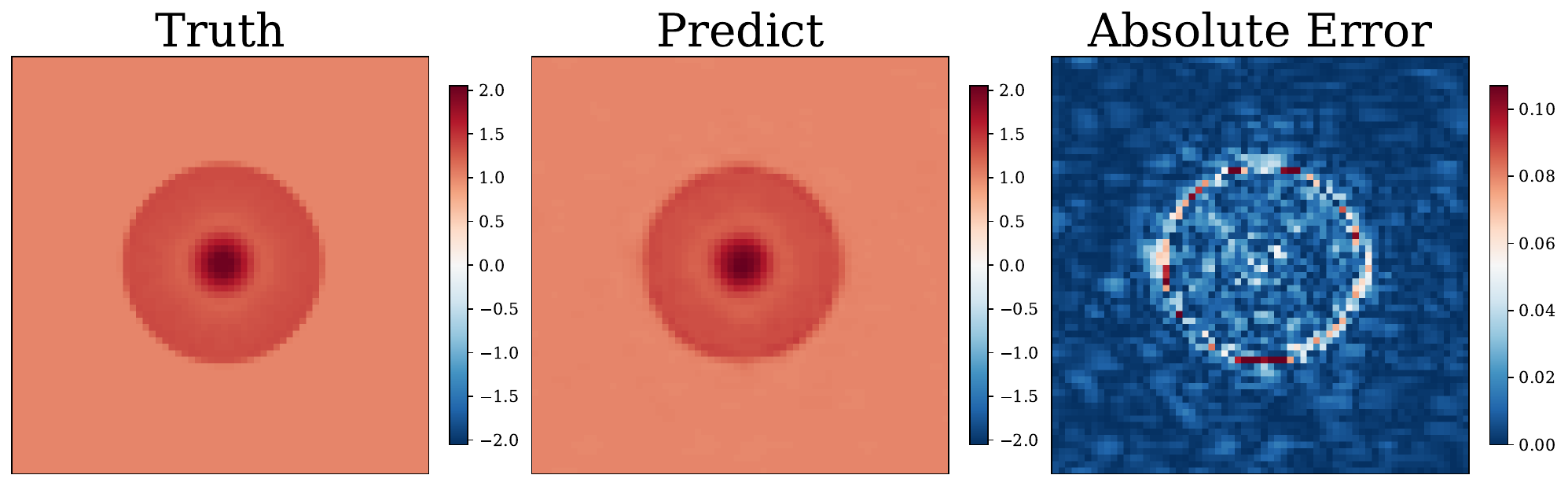}\\[-0.35em]
{\footnotesize\textnormal{(e)}}
\caption{Water-level predictions for the shallow-water equations. Each subfigure compares the ground-truth field (left), the predicted field (center), and the pointwise absolute error (right): (a) CoKo-UNO, (b) FNO2d, (c) KNO2d, (d) RNO, and (e) \NEW{UNO T-channel}.}
\label{fig:app_swe_fields}
\end{figure}

Figs.~\ref{fig:app_kolmogorov_fields}--\ref{fig:app_swe_fields} compare the predicted fields of CoKo-UNO with those of the FNO2d, KNO2d, RNO, and UNO baselines on Kolmogorov flow, cylinder flow, and the shallow-water equations at representative rollout times, together with the corresponding pointwise error maps. At the later time, the baseline predictions exhibit visibly smeared small-scale structures, phase drift relative to the reference vortex street in the cylinder wake, and distorted wave fronts in the shallow-water case, whereas the CoKo-UNO prediction retains coherent structures with errors confined to thin filaments, shear layers, and fronts. These visualizations complement the quantitative diagnostics of Section~\ref{sec:diagnostics}.

\section{Implementation details and reproducibility}
\label{app:implementation}

\subsection{Model configuration}
Table~\ref{tab:model_config} summarizes the configuration of CoKo-UNO used in the $64\times64$ main experiments; the $32\times32$ ablation runs use the same configuration with full compensation where indicated.

\begin{table}[H]
\centering
\caption{Configuration of CoKo-UNO in the $64\times64$ main experiments.}
\label{tab:model_config}
\small
\renewcommand{\arraystretch}{1.08}
\begin{tabular}{@{}>{\raggedright\arraybackslash}p{0.45\linewidth}@{\hspace{1.2em}}>{\raggedright\arraybackslash}p{0.517\linewidth}@{}}
\toprule
\textbf{Component} & \textbf{Setting} \\
\midrule
Backbone levels $L$ & $7$ (three encoder levels, one bottleneck level, and three decoder levels) \\
\addlinespace[2pt]
Retained Fourier modes per level & $12\times12$ \\
\addlinespace[2pt]
Channel widths per level & $32$, $64$, $128$, $128$, $64$, $32$, $16$ \\
\addlinespace[2pt]
Bottleneck latent dimension $d$ & $32$ \\
\addlinespace[2pt]
Selective-SSM state dimension & $16$ \\
\addlinespace[2pt]
Selective-SSM expansion factor / convolution kernel & $2$ / $4$ \\
\addlinespace[2pt]
Compensation strategy & Mamba-based temporal modeling at the bottleneck and low-resolution skip connections; Koopman linear compensation at the remaining skip connections \\
\addlinespace[2pt]
\NEW{Input/output windows $(T_{\mathrm{in}},T_{\mathrm{out}})$} & \NEW{NS $\nu=10^{-3}$: $(10,40)$; NS $\nu=10^{-4}$: $(10,20)$; cylinder: $(10,10)$; SWE: $(10,10)$} \\
\addlinespace[2pt]
Warmup overlap / rollout stride & $5$ / $5$ \\
\addlinespace[2pt]
Total trainable parameters & $4.64\times10^6$ \\
\bottomrule
\end{tabular}
\end{table}

\subsection{Training setup and hardware environment}
All models are trained for $200$ epochs using Adam with an initial learning rate of $10^{-3}$ and a weight decay of $10^{-4}$. We use a StepLR scheduler with a multiplicative factor of $0.5$: the learning rate is decayed every $50$ epochs for FNO2d and KNO2d, and every $80$ epochs for RNO and the temporal-batch UNO variants. At $64\times64$ resolution, FNO2d, KNO2d, and UNO T-channel are trained with a mini-batch size of $100$, whereas RNO and the temporal-batch variants use a mini-batch size of $10$. For the latter, each batch contains $T_{\mathrm{in}}=10$ input time steps, corresponding to an effective batch size of $100$ time-step samples. \NEW{To quantify run-to-run variability consistently, every model on every dataset is trained with the same three random seeds: $42$, $2025$, and $3407$; reported accuracy statistics are the mean $\pm$ standard deviation across these independent runs.}

Experiments are performed using a single NVIDIA A800 80GB PCIe GPU on a server equipped with two Intel Xeon Platinum 8358 processors (64 CPU cores in total) and running Rocky Linux 8.7. The software stack comprises Python 3.10, PyTorch 2.6.0 built with CUDA 12.6, CUDA toolkit 12.6, and cuDNN 9.5.1. All wall-clock measurements reported in this paper, including the per-epoch training times in Table~\ref{tab:efficiency} and the inference times in Table~\ref{tab:complexity}, are collected on this configuration under identical software conditions, ensuring a consistent timing comparison across models.

\subsection{Baseline implementations}
\NEW{All baselines are built from the official public implementations released by their authors, used without architectural modification; only the data-loading and training loops are aligned with the unified protocol of Section~}\ref{sec:setup}\NEW{. FNO2d is implemented with the }\texttt{neuraloperator}\NEW{ library }\cite{li2021fourier}\NEW{, available at }\url{https://github.com/neuraloperator/neuraloperator}\NEW{. KNO2d uses KoopmanLab }\cite{xiong2024koopman}\NEW{, available at }\url{https://github.com/Koopman-Laboratory/KoopmanLab}\NEW{, and RNO uses the official tipping-point forecasting implementation }\cite{liu2026tipping}\NEW{, available at }\url{https://github.com/neuraloperator/tipping-point-forecast}\NEW{. The UNO variants follow the original UNO implementation }\cite{rahman2023uno}\NEW{, available at }\url{https://github.com/ashiq24/UNO}\NEW{. Using the authors' own implementations, rather than third-party reimplementations, ensures that each baseline is evaluated at its reported capability and that the comparison does not disadvantage any method.}

\subsection{Data availability}
\NEW{The Kolmogorov-flow data are downloaded from the official NeuralOperator library, following the public FNO benchmark configuration}~\cite{li2021fourier}. \NEW{The underlying trajectories are generated on a uniform $256\times256$ grid using a pseudo-spectral split-step solver: the nonlinear and forcing terms are advanced by Heun's method, whereas the viscous term is advanced by a Crank--Nicolson update. The internal solver time step is $10^{-4}$; fields are saved at one-unit physical-time intervals and downsampled to $64\times64$. The total simulated physical times are $T=50$ for $\nu=10^{-3}$ and $T=30$ for $\nu=10^{-4}$.} The cylinder-flow data are taken from CFDBench \cite{luo2023cfdbench}, while the shallow-water data are taken from PDEBench \cite{takamoto2022pdebench}. All datasets use $1{,}000$ training and $200$ test trajectories, as described in Section~\ref{sec:setup}. \NEW{The official NeuralOperator repository is available at }\url{https://github.com/neuraloperator/neuraloperator}\NEW{. The CFDBench data and associated utilities are available at }\url{https://github.com/luo-yining/CFDBench}\NEW{. The PDEBench shallow-water data are archived at }\url{https://doi.org/10.18419/darus-2986}\NEW{, with accompanying download and generation utilities available at }\url{https://github.com/pdebench/PDEBench}\NEW{.}

\subsection{Code availability}
\NEW{The complete training and evaluation code of CoKo-UNO, together with the configuration files and scripts needed to reproduce every table and figure in this paper, will be released at }\url{https://github.com/tangyinglv/CoKo-UNO}\NEW{ upon publication.}
\bibliographystyle{elsarticle-num}
\bibliography{CoKo_UNO_references_v3}

@article{raissi2019physics,author={Raissi, Maziar and Perdikaris, Paris and Karniadakis, George Em},title={Physics-informed neural networks: A deep learning framework for solving forward and inverse problems involving nonlinear partial differential equations},journal={Journal of Computational Physics},volume={378},pages={686--707},year={2019}}

@article{karniadakis2021physics,author={Karniadakis, George Em and Kevrekidis, Ioannis G and Lu, Lu and Perdikaris, Paris and Wang, Sifan and Yang, Liu},title={Physics-informed machine learning},journal={Nature Reviews Physics},volume={3},pages={422--440},year={2021}}

@article{lu2021learning,author={Lu, Lu and Jin, Pengzhan and Karniadakis, George Em},title={Learning nonlinear operators via {DeepONet} based on the universal approximation theorem of operators},journal={Nature Machine Intelligence},volume={3},pages={218--229},year={2021}}

@article{kovachki2023neural,author={Kovachki, Nikola and others},title={Neural operator: Learning maps between function spaces},journal={Journal of Machine Learning Research},volume={24},number={89},pages={1--97},year={2023}}

@inproceedings{li2021fourier,author={Li, Zongyi and Kovachki, Nikola and Azizzadenesheli, Kamyar and others},title={Fourier Neural Operator for Parametric Partial Differential Equations},booktitle={International Conference on Learning Representations},year={2021}}

@article{rahman2023uno,author={Rahman, Md Ashiqur and Ross, Zachary E and Azizzadenesheli, Kamyar},title={U-{NO}: U-shaped Neural Operators},journal={Transactions on Machine Learning Research},year={2023}}

@article{li2022geometry,author={Li, Zongyi and Huang, Daniel Zhengyu and Liu-Schiaffini, Michele and others},title={Fourier neural operator with learned deformations for {PDE}s on general geometries},journal={Journal of Machine Learning Research},year={2023}}

@inproceedings{li2023gino,author={Li, Zongyi and Huang, Daniel Zhengyu and Liu-Schiaffini, Michele and others},title={Geometry-informed Neural Operator for Large-scale 3D {PDE}s},booktitle={Advances in Neural Information Processing Systems},year={2023}}

@inproceedings{hao2023gnot,author={Hao, Zhongkai and Ying, Chen and Su, Zhengyi and others},title={G{N}O: A General Neural Operator Transformer for Operator Learning},booktitle={International Conference on Machine Learning},year={2023}}

@inproceedings{cao2021choose,author={Cao, Shuhao},title={Choose a Transformer: Fourier or Galerkin},booktitle={Advances in Neural Information Processing Systems},year={2021}}

@article{koopman1931,author={Koopman, Bernard O},title={Hamiltonian systems and transformation in Hilbert space},journal={Proceedings of the National Academy of Sciences},volume={17},number={5},pages={315--318},year={1931}}

@article{mezic2005spectral,author={Mezi{\'c}, Igor},title={Spectral properties of dynamical systems, model reduction and decompositions},journal={Nonlinear Dynamics},volume={41},pages={309--325},year={2005}}

@article{mezic2013analysis,author={Mezi{\'c}, Igor},title={Analysis of fluid flows via spectral properties of the {K}oopman operator},journal={Annual Review of Fluid Mechanics},volume={45},pages={357--378},year={2013}}

@article{schmid2010dynamic,author={Schmid, Peter J},title={Dynamic mode decomposition of numerical and experimental data},journal={Journal of Fluid Mechanics},volume={656},pages={5--28},year={2010}}

@article{williams2015data,author={Williams, Matthew O and Kevrekidis, Ioannis G and Rowley, Clarence W},title={A data-driven approximation of the Koopman operator: Extending dynamic mode decomposition},journal={Journal of Nonlinear Science},volume={25},pages={1307--1346},year={2015}}

@article{lusch2018deep,author={Lusch, Bethany and Kutz, J Nathan and Brunton, Steven L},title={Deep learning for universal linear embeddings of nonlinear dynamics},journal={Nature Communications},volume={9},number={4950},year={2018}}

@inproceedings{takeishi2017learning,author={Takeishi, Naoya and Kawahara, Yoshinobu and Yairi, Takehisa},title={Learning Koopman invariant subspaces for dynamic mode decomposition},booktitle={Advances in Neural Information Processing Systems},year={2017}}

@article{korda2018linear,author={Korda, Milan and Mezi{\'c}, Igor},title={Linear predictors for nonlinear dynamical systems: Koopman operator meets model predictive control},journal={Automatica},volume={93},pages={149--160},year={2018}}

@article{xiong2024koopman,author={Xiong, Wei and Huang, Xiaomeng and Zhang, Ziyang and Deng, Ruixuan and Sun, Pei and Tian, Yang},title={Koopman neural operator as a mesh-free solver of non-linear partial differential equations},journal={Journal of Computational Physics},volume={513},pages={113194},year={2024}}

@article{kalman1960new,author={Kalman, Rudolf E},title={A New Approach to Linear Filtering and Prediction Problems},journal={Journal of Basic Engineering},volume={82},number={1},pages={35--45},year={1960}}

@inproceedings{gu2022s4,author={Gu, Albert and Goel, Karan and R{\'e}, Christopher},title={Efficiently Modeling Long Sequences with Structured State Spaces},booktitle={International Conference on Learning Representations},year={2022}}

@inproceedings{gu2023mamba,author={Gu, Albert and Dao, Tri},title={Mamba: Linear-Time Sequence Modeling with Selective State Spaces},booktitle={First Conference on Language Modeling},year={2024}}

@inproceedings{bengio2015scheduled,author={Bengio, Samy and Vinyals, Oriol and Jaitly, Navdeep and Shazeer, Noam},title={Scheduled Sampling for Sequence Prediction with Recurrent Neural Networks},booktitle={Advances in Neural Information Processing Systems},year={2015}}

@inproceedings{lippe2023pde,author={Lippe, Phillip and Veeling, Bastiaan S and Perdikaris, Paris and Turner, Richard E and Brandstetter, Johannes},title={{PDE}-Refiner: Achieving Accurate Long Rollouts with Neural {PDE} Solvers},booktitle={Advances in Neural Information Processing Systems},volume={36},year={2023}}

@inproceedings{takamoto2022pdebench,author={Takamoto, Makoto and Praditia, Tommy and Leiteritz, Raphael and MacKinlay, Daniel and Alesiani, Francesco and Pflueger, Dirk and Niepert, Mathias},title={{PDEBench}: An Extensive Benchmark for Scientific Machine Learning},booktitle={Advances in Neural Information Processing Systems},volume={35},year={2022}}

@article{luo2023cfdbench,author={Luo, Yining and Chen, Yingfa and Zhang, Zhen},title={{CFDBench}: A Large-Scale Benchmark for Machine Learning Methods in Fluid Dynamics},journal={arXiv preprint arXiv:2310.05963},year={2023}}

@inproceedings{brandstetter2022message,author={Brandstetter, Johannes and Worrall, Daniel and Welling, Max},title={Message Passing Neural {PDE} Solvers},booktitle={International Conference on Learning Representations},year={2022}}

@inproceedings{mno2020,author={Li, Zongyi and Kovachki, Nikola and Azizzadenesheli, Kamyar and Liu, Burigede and Stuart, Andrew and Bhattacharya, Kaushik and Anandkumar, Anima},title={Multipole Graph Neural Operator for Parametric Partial Differential Equations},booktitle={Advances in Neural Information Processing Systems},volume={33},year={2020}}

@article{pino2024,author={Li, Zongyi and Zheng, Hongkai and Kovachki, Nikola and Jin, David and Chen, Haoxuan and Liu, Burigede and Azizzadenesheli, Kamyar and Anandkumar, Anima},title={Physics-Informed Neural Operator for Learning Partial Differential Equations},journal={ACM/IMS Journal of Data Science},volume={1},pages={1--27},year={2024}}

@inproceedings{raonic2023cno,author={Raoni{\'c}, Bogdan and Molinaro, Roberto and De Ryck, Tim and Rohner, Tobias and Bartolucci, Francesca and Alaifari, Rima and Mishra, Siddhartha and de B{\'e}zenac, Emmanuel},title={Convolutional Neural Operators for robust and accurate learning of {PDE}s},booktitle={International Conference on Learning Representations},volume={36},year={2023}}

@inproceedings{pfaff2021learning,author={Pfaff, Tobias and Fortunato, Meire and Sanchez-Gonzalez, Alvaro and Battaglia, Peter},title={Learning Mesh-Based Simulation with Graph Networks},booktitle={International Conference on Learning Representations},year={2021}}

@inproceedings{furuya2024continuous,author={Furuya, Takashi and Puthawala, Michael Anthony and Lassas, Matti and de Hoop, Maarten V},title={Can neural operators always be continuously discretized?},booktitle={Advances in Neural Information Processing Systems},year={2024}}

@inproceedings{wang2024latent,author={Wang, Tian and Wang, Chuang},title={Latent Neural Operator for Solving Forward and Inverse {PDE} Problems},booktitle={Advances in Neural Information Processing Systems},year={2024}}

@inproceedings{rahman2024pretraining,author={Rahman, Md Ashiqur and others},title={Pretraining Codomain Attention Neural Operators for Solving Multiphysics {PDE}s},booktitle={Advances in Neural Information Processing Systems},year={2024}}

@inproceedings{zheng2024aliasfree,author={Zheng, Jianwei and Luo, Liwei and Xu, Ni and Zhu, Junwei and Lin, Xiaoxu and Zhang, Xiaoqin},title={Alias-Free Mamba Neural Operator},booktitle={Advances in Neural Information Processing Systems},year={2024}}

@inproceedings{cao2025spectralrefiner,author={Cao, Shuhao and Brarda, Francesco and Li, Rui Peng and Xi, Yuanzhe},title={Spectral-Refiner: Accurate Fine-Tuning of Spatiotemporal Fourier Neural Operator for Turbulent Flows},booktitle={International Conference on Learning Representations},year={2025}}

@inproceedings{song2026adaptive,author={Song, Zeyuan and Jiang, Zheyu},title={Adaptive Mamba Neural Operators},booktitle={International Conference on Learning Representations},year={2026}}

@book{pope2000turbulent,author={Pope, Stephen B},title={Turbulent Flows},publisher={Cambridge University Press},year={2000}}

@book{leveque2002finite,author={LeVeque, Randall J},title={Finite Volume Methods for Hyperbolic Problems},publisher={Cambridge University Press},year={2002}}

@article{williamson1996vortex,author={Williamson, Charles H K},title={Vortex dynamics in the cylinder wake},journal={Annual Review of Fluid Mechanics},volume={28},pages={477--539},year={1996}}

@book{courant1962methods,author={Courant, Richard and Hilbert, David},title={Methods of Mathematical Physics, Volume {II}: Partial Differential Equations},publisher={Interscience Publishers},year={1962}}

@article{harten1987uniformly,author={Harten, Amiram and Engquist, Bj{\"o}rn and Osher, Stanley and Chakravarthy, Sukumar R},title={Uniformly high order accurate essentially non-oscillatory schemes, {III}},journal={Journal of Computational Physics},volume={71},number={2},pages={231--303},year={1987}}

@article{shu1988efficient,author={Shu, Chi-Wang and Osher, Stanley},title={Efficient implementation of essentially non-oscillatory shock-capturing schemes},journal={Journal of Computational Physics},volume={77},number={2},pages={439--471},year={1988}}

@article{geneva2020modeling,author={Geneva, Nicholas and Zabaras, Nicholas},title={Modeling the dynamics of {PDE} systems with physics-constrained deep auto-regressive networks},journal={Journal of Computational Physics},volume={403},pages={109056},year={2020}}

@article{chandler2013invariant,author={Chandler, Gary J and Kerswell, Richard R},title={Invariant recurrent solutions embedded in a turbulent two-dimensional {Kolmogorov} flow},journal={Journal of Fluid Mechanics},volume={722},pages={554--595},year={2013}}

@article{liu2026tipping,author={Miguel Liu-Schiaffini and Clare E. Singer and Nikola Kovachki and Sze Chai Leung and Hyunji Jane Bae and Kamyar Azizzadenesheli and Anima Anandkumar},title={Tipping Point Forecasting in Non-Stationary Dynamics on Function Spaces},journal={arXiv preprint arXiv:2308.08794},year={2026}}

\end{document}